\documentclass{emulateapj}
\usepackage{psfig,natbib,amsmath}

\newcommand{\citepeg}[1]{\citep[{e.g.,}][]{#1}}
\newcommand{\mnhi}{N_{\rm HI}}
\begin{document}

\title{GRB 071003: Broadband Follow-up Observations of a Very Bright Gamma-Ray Burst in a Galactic Halo}

\def\berk{1}
\def\lick{2}
\def\uva{3}
\def\jansky{4}
\def\sloan{5}
\def\mich{6}
\def\davis{7}
\def\caltech{8}
\def\chicago{9}
\def\nrao{10}
\def\keck{11}
\def\arizona{12}
\def\aerospace{13} 
\def\boeing{14}
\def\rice{15}
\def\benedict{16}

\author{D.~A.~Perley\altaffilmark{\berk},
        W. Li\altaffilmark{\berk},
        R.~Chornock\altaffilmark{\berk},
        J.~X.~Prochaska\altaffilmark{\lick},
        N.~R.~Butler\altaffilmark{\berk},
        P.~Chandra\altaffilmark{\uva,\jansky},
        L.~K.~Pollack\altaffilmark{\lick},
        J.~S.~Bloom\altaffilmark{\berk,\sloan},		
        A.~V.~Filippenko\altaffilmark{\berk},
        H.~Swan\altaffilmark{\mich},
        F.~Yuan\altaffilmark{\mich}
        C.~Akerlof\altaffilmark{\mich},
        M.~W.~Auger\altaffilmark{\davis},
        S.~B.~Cenko\altaffilmark{\caltech},
        H.-W.~Chen\altaffilmark{\chicago},
        C.~D.~Fassnacht\altaffilmark{\davis},
        D.~Fox\altaffilmark{\caltech},
        D.~Frail\altaffilmark{\nrao},
        E.~M.~Johansson\altaffilmark{\keck},
        D.~Le~Mignant\altaffilmark{\lick,\keck},
        T.~McKay\altaffilmark{\mich},
        M.~Modjaz\altaffilmark{\berk},
        W.~Rujopakarn\altaffilmark{\arizona},
        R.~Russell\altaffilmark{\aerospace},
        M.~A.~Skinner\altaffilmark{\boeing},
        G.~H.~Smith\altaffilmark{\lick},
        I.~Smith\altaffilmark{\rice},
        M.~A.~van~Dam\altaffilmark{\keck}, and
        S.~Yost\altaffilmark{\benedict}
        }
 \altaffiltext{\berk}{Department of Astronomy, 
        University of California, Berkeley, CA 94720-3411.}
 \altaffiltext{\lick}{Department of Astronomy and Astrophysics, UCO/Lick Observatory; 
        University of California, 1156 High Street, Santa Cruz, CA 95064.}
 \altaffiltext{\uva}{Department of Astronomy, University of Virginia, P.O. Box 400325,
        Charlottesville, VA 22904.}
 \altaffiltext{\jansky}{Jansky Fellow, National Radio Astronomy Observatory.}
 \altaffiltext{\sloan}{Sloan Research Fellow.}
 \altaffiltext{\mich}{University of Michigan, Randall Laboratory of Physics, 
        450 Church St., Ann Arbor, MI, 48109-1040.}
 \altaffiltext{\davis}{Department of Physics, University of California, 1 Shields Avenue, Davis, CA 95616.}
 \altaffiltext{\caltech}{Division of Physics, Mathematics, and Astronomy, MS 105-24, 
        California Institute of Technology, Pasadena, CA 91125.}
 \altaffiltext{\chicago}{Department of Astronomy and Astrophysics, University of Chicago, 
        5640 S. Ellis Ave, Chicago, IL 60637.}
 \altaffiltext{\nrao}{National Radio Astronomy Observatory, P.O. Box O, Socorro, NM 87801.}
 \altaffiltext{\keck}{W. M. Keck Observatory, 65-1120 Mamalahoa Highway, Kamuela, HI 96743.}
 \altaffiltext{\arizona}{Steward Observatory, Tucson, AZ 85721.}
 \altaffiltext{\aerospace}{The Aerospace Corporation, Mail Stop M2-266, PO Box 92957, Los Angeles, CA 90009-29957.}
 \altaffiltext{\boeing}{The Boeing Company, 535 Lipoa Parkway, Suite 200, Kihei, HI 96753.}
 \altaffiltext{\rice}{Department of Physics and Astronomy, Rice University, 6100 South Main, MS108, Houston, TX 77251-1892.}
 \altaffiltext{\benedict}{College of St. Benedict, St. Joseph, MN 56374.}
\email{(dperley,wli)@astro.berkeley.edu}

\slugcomment{To appear in ApJ 2008 November 10}

\begin{abstract}
The optical afterglow of long-duration GRB 071003 is
among the brightest yet to be detected from any GRB, with $R \approx
12$ mag in KAIT observations starting
42~s after the GRB trigger, including filtered detections during
prompt emission.  However, our high S/N ratio afterglow
spectrum displays only extremely weak absorption lines at what we
argue is the host redshift of $z=1.60435$ --- in contrast to the three
other, much stronger \ion{Mg}{2} absorption systems observed at lower
redshifts.  Together with Keck adaptive optics observations which fail
to reveal a host galaxy coincident with the burst position, our
observations suggest a halo progenitor and offer a cautionary tale
about the use of \ion{Mg}{2} for GRB redshift determination.  We
present early through late-time observations spanning the
electromagnetic spectrum, constrain the connection between the prompt
emission and early variations in the light curve (we observe no
correlation), and discuss possible origins for an unusual, marked
rebrightening that occurs a few hours after the burst: likely either a
late-time refreshed shock or a wide-angle secondary jet.  Analysis of
the late-time afterglow is most consistent with a wind environment,
suggesting a massive star progenitor.  Together with GRB~070125, this
may indicate that a small but significant portion of star formation in
the early universe occurred far outside what we consider a normal
galactic disk.
\end{abstract}


\section{Introduction}

Concurrent observations of long-wavelength afterglow and ongoing
gamma-ray burst (GRB) activity should, in principle, yield important
constraints on the nature of the physical processes of the emission
(eg., \citealt{Kobayashi2000}). However, as a GRB typically lasts less
than 100~s, it is challenging for large ground-based optical/infrared
follow-up facilities to react to a GRB alert quickly and take data
during the prompt phase. Multi-color observations, which provide vital
information on the emission mechanism, are even more difficult to
obtain during the prompt phase because of the added overhead
associated with changing filters. Nevertheless, due to the coordinated
efforts of recent space missions (\textit{HETE-II}, \citealt{Ricker+2003};
\textit{Swift}, \citealt{Gehrels+2004}) to detect GRBs and various
ground-based optical follow-up programs, observations during the prompt
phase of GRBs are no longer uncommon --- the optical afterglows (OAs)
of several dozen GRBs have been observed
\citepeg{Akerlof+1999,Vestrand+2006,Yost+2007} during gamma-ray
emission, and multi-color optical data have been obtained in a handful
of cases \citepeg{Blake+2005,Nysewander+2007}.

Observations of GRBs in the past several years have also revealed
a rich demography in OA behavior.  Some OAs have monotonic power-law
decays \citepeg{Li+2003a,Laursen+2003}, while others have plateau
\citepeg{Rykoff+2006} and rebrightening \citepeg{Wozniak+2006}
phases.  Even among GRBs with relatively simple behavior, however,
short-timescale features not predicted in the basic shock models often
appear in sufficiently well-sampled data.  Various modifications to
the standard picture have been proposed to explain such observations,
including the presence of a jet with single (e.g., \citealt{Sari+1999}) 
or multiple (e.g., \citealt{Berger+2003}) components, refreshed shocks
\citep{Zhang+2006}, central engine activity
\citep{Kocevski+2007,Chincarini+2007}, gravitational microlensing
\citep{Garnavich+2000}, and density irregularity in the GRB
environment \citep{Holland+2003}.  Observationally, constraints on the
change in the afterglow color and the spectral energy distribution
(SED) play an important role in limiting the viability of models for a
particular GRB.

The question of the nature of the GRB itself is intimately tied to the
question of its environment and origins.  At intermediate to late
times, spectroscopy of the afterglow
\citepeg{Prochaska+2007,DElia+2007} and deep imaging of the host
environment \citep[e.g.,][]{Bloom+2002, Fruchter+2006} can help
establish the nature of the GRB's progenitor and environment,
connecting what we learn about the burst itself to the larger question
of its origins and place in the early universe.

In this paper, we report on our photometric and spectroscopic
observations of GRB\,071003 with various telescopes from the prompt
phase to late times.  In \S 2 we describe the observations, and in \S
3 we present the reductions.  The analysis of the light curves and
the constraints on the changes in the colors and SEDs are given in \S
4. The conclusions, including the implications of the extremely
unusual spectrum of this event, are discussed in \S 5.  We assume
$H_{\circ}=71$ km s$^{-1}$ Mpc$^{-1}$, $\Omega_{\rm M}=0.3$, and
$\Omega_{\Lambda}=0.7$ throughout the paper.  These reports should be
considered the final analysis of our group's data on GRB\,071003,
superseding any previously announced results (e.g., in the GCN
Circulars).

\section{Observations}

\subsection{BAT/XRT Observations} 

On 2007 October 3, 07:40:55 UT (defined as $t=0$ in this paper; UT dates
are used throughout), a bright GRB triggered the Burst Alert Telescope (BAT)
onboard the \textit{Swift} satellite (trigger 292934
\citealt{GCN6837}).  The first GCN notice was distributed within 16~s.
Unfortunately, \textit{Swift} was still returning to normal
observations after its 2007 August gyro failure, but it did slew to the
position after 22~ks and began observations using the X-Ray Telescope (XRT).

We downloaded the \textit{Swift} BAT and XRT data from the {\it
Swift}~Archive\footnote{ftp://legacy.gsfc.nasa.gov/swift/data } and
quicklook data site.\footnote{\url
http://swift.gsfc.nasa.gov/cgi-bin/sdc/ql }  The XRT and BAT spectra
were fitted using ISIS\footnote{\url http://space.mit.edu/CXC/ISIS }.

The XRT data were processed with version 0.11.4 of the {\tt
xrtpipeline} reduction script from the HEAsoft~6.3.1\footnote{\url
http://heasarc.gsfc.nasa.gov/docs/software/lheasoft/ } software
release.  We employ the latest (2007 December 4) XRT calibration
files.  Our reduction of XRT data from cleaned event lists output by
{\tt xrtpipeline} to science-ready light curves and spectra is
described in detail by \cite{ButlerKocevski2007a}.  We use the latest
calibration files from the 2007 September 24 BAT database release.  We
establish the energy scale and mask weighting for the BAT event mode
data by running the {\tt bateconvert} and {\tt batmaskwtevt} tasks.
Spectra and light curves are extracted with the {\tt batbinevt} task,
and response matrices are produced by running {\tt batdrmgen}.  
To produce the BAT spectra, we
apply the systematic error corrections to the low-energy BAT spectral
data as suggested by the BAT Digest Web site\footnote{\url
http://swift.gsfc.nasa.gov/docs/swift/analysis/bat\_digest.html}, and
fit the data in the 15--150 keV band.  The spectral normalizations are
corrected for satellite slews using the {\tt batupdatephakw} task.

\begin{figure}
\centerline{\includegraphics[width=3.5in,angle=0]{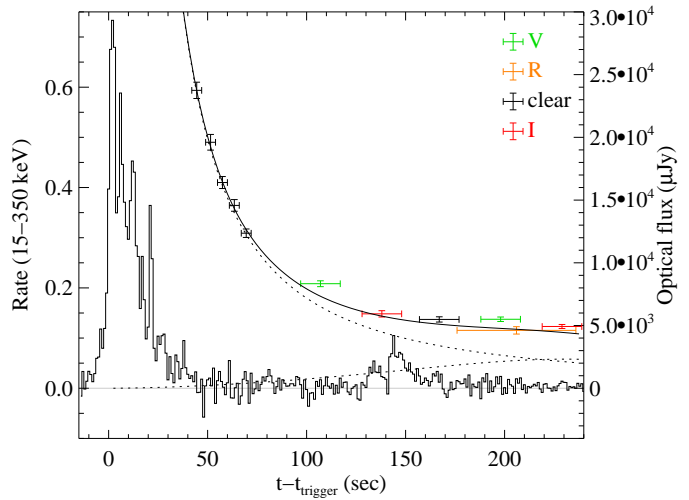}}
\caption[relation] {Light curve from the \textit{Swift} BAT of GRB
071003, with optical photometry from KAIT and P60, and the optical
light curve model discussed in Section 4.3, overplotted.  The GRB
is dominated by a complicated, spiky emission episode
in the first 30~s, but a pulse is also observed much later, at 150~s.
Optical data points (all from KAIT, except one $R$-band measurement from
the P60), by contrast, show a power-law decay at early times followed by a
slow-rising ``bump.''  Here the $V$ and $I$ filtered observations have
been offset to match the $R$ and unfiltered points based on the
relative colors at 2000~s.}
\label{fig:grblc}
\end{figure}

The burst exhibits one dominant emission episode of duration $dt\approx
30$~s, followed by a minor pulse $\sim 150$~s later of duration $\sim
20$~s.  The total duration is $T_{90}=148\pm1$~s,
\footnote{All uncertainties quoted in this paper are 1$\sigma$, except 
where specified otherwise.}
placing it clearly into the long GRB class.
The primary pulse is resolved into multiple pulses.
 The gamma-ray light curve is shown in Figure \ref{fig:grblc},
overplotted with early-time photometry from KAIT and P60
(discussed in \S 3.2 and \S 3.5, respectively).


The time-integrated BAT spectrum from $t=-10.3$ to $t=169$~s is
acceptably fitted ($\chi^2/\nu=47.64/55$, where $\nu$ is the number of
degrees of freedom) by a power-law model, with
photon index $\alpha = -1.3\pm 0.1$ and energy fluence
$S_{\gamma}=(1.7\pm 0.1) \times 10^{-5}$ erg cm$^{-2}$ (15--350 keV).
The main emission episode ($t=-1.4$~s to $t=22.8$~s) is harder
($\alpha = -1.08\pm 0.03$, $S_{\gamma}=(1.51\pm0.03) \times 10^{-5}$
erg cm$^{-2}$, $\chi^2/\nu=56.71/55$), while the final pulse
($t=131$--$169$~s) is softer ($\alpha = -1.8\pm 0.2$,
$S_{\gamma}=1.2^{+0.1}_{-0.2} \times 10^{-6}$ erg cm$^{-2}$,
$\chi^2/\nu=41.15/55$).

X-ray observations with the XRT began 6.2 hr after the BAT trigger.
The X-ray light curve measured until $t\approx 5\times 10^{5}$~s is
well fitted by a power-law time decay $t^{-1.68\pm0.03}$.  The 
time-integrated spectrum is well fitted ($\chi^2/\nu=48.47/54$) by an absorbed
power-law model [photon index $\Gamma=2.14\pm 0.12$, unabsorbed
$F_X=(5.8\pm 0.4) \times 10^{-13}$ erg cm$^{-2}$ s$^{-1}$].  The
equivalent H column density, $N_{\rm H} = (2.2\pm 0.4) \times 10^{21}$
cm$^{-2}$, is marginally consistent with the expected Galactic column
density in the source direction, $N_{\rm H}=1.1\times 10^{21}$ cm$^{-2}$
\citep{Dickey1990}.  Examining the X-ray hardness ratio
\citep[e.g.,][]{ButlerKocevski2007b}, there is no evidence for
spectral evolution during the XRT observation.

\subsection{KAIT Observations} 

The Katzman Automatic Imaging Telescope (KAIT) is a 0.76-m robotic
telescope at Lick Observatory that is dedicated to searching for and
observing supernovae and monitoring other variable or ephemeral
celestial phenomena. It is equipped with a Finger Lakes Instrument
(FLI) ProLine PL77 back-illuminated CCD camera having a resolution of
$0\farcs8$ pixel$^{-1}$ and a total field of view (FOV) of $\sim$
$6\farcm8\times6\farcm8$.  More information on KAIT can be found in
\citet{Li+2000}, \citet{Filippenko+2001}, and \citet{Filippenko2005},
while the KAIT GRB alert system is described in detail by
\citet{Li+2003a}.  Notable KAIT observations of GRBs include
GRB~021211 \citep{Li+2003b}, GRB~051111 \citep{Butler+2006},
GRB~060210 (\citealt{GCN4727}, Li et al. 2008 in preparation), and
GRB~080319B \citep{Bloom+2008}.

Several improvements have been implemented for the KAIT GRB alert
system since the description given by \citet{Li+2003a}. An FLI PL77
camera has replaced the Apogee AP7 camera, offering a much faster
readout time (1.2~s for FLI vs. 11.0~s for Apogee). A new feature
has been incorporated into the software so the system can easily
terminate an ongoing exposure in preparation for the GRB response
sequence. Most importantly, a real-time image-processing pipeline has
been developed to compare the KAIT images to archival Digital Sky
Survey (DSS) images to identify new objects. Astrometry solutions are
derived for the KAIT images by matching the detected objects to the
USNO B1 catalog \citep{Monet+2003}, providing coordinates to any new
objects to a precision of $\sim$0$\arcsec$.2. Point-spread-function
(PSF) fitting photometry is also performed on new objects, and
calibrated to the red magnitudes of the stars in the USNO B1 catalog.
The image-processing results are displayed in real time on a
website.\footnote{{\url
http://hercules.berkeley.edu/grbdata/grbfinder.gif .}}

\begin{figure*}
\centerline{\includegraphics[width=5.5in,angle=0]{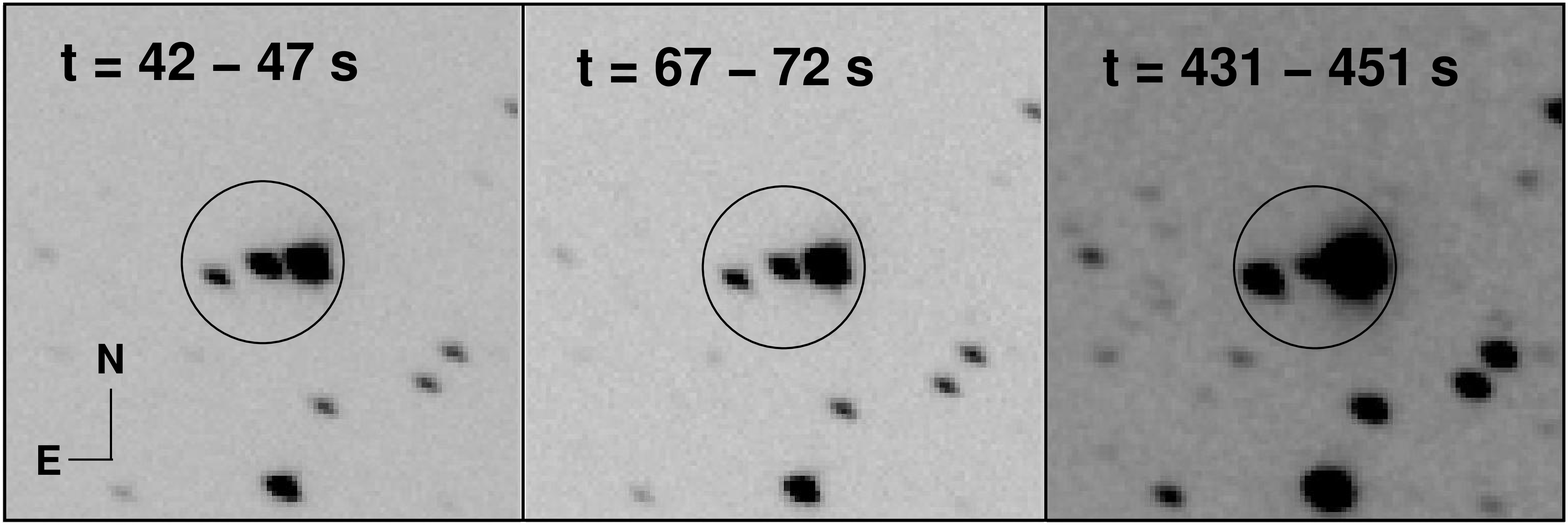}}
\caption[relation] {Sequence of KAIT images for the OA of GRB
071003.  An $80\arcsec \times 80\arcsec$ section is shown for the
first and fifth unfiltered 5~s images and for a 20~s unfiltered image
that started at $t = 431$~s. The OA is the central object in the
circles. It is well detected in the early images and rapidly fades.
The image quality is poor owing to the very high airmass of the object.}
\label{fig:kaitseq}
\end{figure*}

For GRB\,071003, the KAIT GRB alert program received the GCN socket
notice at $t = 16$~s. The system immediately terminated the ongoing
supernova search program and began to slew the telescope to the GRB
position.  After slewing from close to meridian to an hour angle of
4.2 hr, a sequence of 5$\times$5~s unfiltered images began at
$t = 42$~s.  KAIT then switched to a sequence that alternated with
20~s $V$, $I$, and unfiltered images.  Finally, the sequence
converted to 20~s $I$ and unfiltered images. Because of the physical
west hour angle limit of 4.7 hr, KAIT only finished part of this
pre-arranged sequence. In total, 56 images were obtained in the $V$,
$I$, and unfiltered passbands from $t = 42$ to $1628$~s, with full
width at half-maximum intensity (FWHM) of $\sim$3$\arcsec$.

Visual inspection of the image-processing results revealed a true new
object, first reported by our group \citep{GCN6838}, measured at 12.8
mag at a position of $\alpha$ = $20^h07^m24^s.12$, $\delta$ =
$+10^\circ56\arcmin51\arcsec.8$ (equinox 2000.0; approximate $1\sigma$
astrometric uncertainty 0\arcsec.3).  Our candidate OA was
subsequently confirmed by observations from the automated Palomar
60-inch (1.5 m) telescope (P60; \citealt{GCN6839}). Further preliminary analysis of OA
early-time behavior from the KAIT observations was reported by
\citet{GCN6844}. Figure \ref{fig:kaitseq} shows a sequence of the KAIT
images for the OA of GRB\,071003.  An $80\arcsec \times 80\arcsec$
section is shown for the first and fifth unfiltered 5~s image and a
20~s unfiltered image that started at $t = 431$~s. As seen in Figure
\ref{fig:kaitseq} and reported by several groups
\citep{GCN6838,GCN6839,GCN6840,GCN6844}, a bright ($R \approx 11$ mag)
foreground star is located 6$\arcsec$.5 west of the OA of GRB
071003. As discussed in \S 3, the presence of this bright star
complicates the photometry for the OA, and various methods have been
used to minimize its contamination.

\subsection{P60 Observations}

The Palomar 60-inch telescope (P60; \citealt{cfm+06}) automatically
responded to the \textit{Swift} trigger for GRB\,071003, beginning a
pre-programmed sequence of observations at 07:43:51 UT (176~s after
the trigger).  Observations were taken in the Kron $R$, Sloan
$i^{\prime}$ and $z^{\prime}$, and Gunn $g$ filters at large airmass
($> 2.5$).  Individual images were reduced in real time by our
automated reduction pipeline.  Manual inspection revealed a fading
point source \citep{GCN6839} in all four filters at the location
reported by \citet{GCN6838}.

A second epoch of observations was manually scheduled for the night of
UT October 4. In an attempt to lessen the contamination of the nearby
bright saturated star, these observations were taken in the Johnson
$V$-band filter in relatively short (30~s) exposures. A sequence of
30 images was obtained.

\subsection{AEOS Observations}

The 3.6-m US Air Force Advanced Electro-Optical System (AEOS)
telescope, located at the Maui Space Surveillance System on Haleakala\footnote{Based on data from the Maui Space Surveillance System, which is operated by Detachment 15 of the U.S. Air Force Research Laboratory's Directed Energy Directorate.},
observed the OA of GRB 071003 with the AEOS Burst Camera (ABC,
\citealt{Swan+2006}). ABC has a back-illuminated $2048 \times 2048$
pixel EEV chip, with a scale of 0$\arcsec$.189 pixel$^{-1}$ and a FOV
of $\sim$ $6\farcm5\times6\farcm5$.  Because there is no direct internet
access to AEOS, after \textit{Swift} detected the GRB, a FAX alert 
was automatically sent to the AEOS control room, to initiate a series 
of Target-of-Opportunity (ToO) observations.  

The AEOS observations of GRB 071003 are all unfiltered 10~s exposures.
The first batch of images started at $\sim$9 minutes after the BAT
trigger, and 238 images were observed until $t \approx 83$ minutes, all
with very good image quality (FWHM $\approx$ 0$\arcsec$.9). The second
batch of images started at $t \approx 205$ minutes, and 56 images were
observed until $t \approx 222$ minutes. Due to the large airmass for these
observations and the degraded seeing conditions, however, the images
have rather poor quality. We have tried various methods to measure the
brightness of the OA in these images but failed. Accordingly, only the
first batch of 238 images is analyzed in this study.  Preliminary
analysis of the AEOS observations is reported by \citet{GCN6841}.

\subsection{Keck I/Gemini-S observations}

In response to the detection of the OA of GRB 071003, we organized a
campaign to obtain spectroscopy and late-time photometry with the 10-m
Keck I and the 8-m Gemini-S telescopes. At $t \approx 2.6$ hr, we
attempted to observe the OA with the HIRES spectrograph at Keck I, but
the data are of poor signal-to-noise ratio (S/N) and no obvious lines
were detected \citep{GCN6843}. Just before the HIRES spectroscopy
started, we also obtained guider images for the OA, providing
important photometric coverage during a gap in the photometry obtained
elsewhere (see \S 3.4). The guider images have a scale of
0$\arcsec$.37 pixel$^{-1}$ with a FOV of 53$\arcsec$.5 $\times$
71$\arcsec$.3.

On 2007 October 4, we observed the GRB\,071003 OA with the Low Resolution
Imaging Spectrometer (LRIS; \citet{Oke+1995}) on Keck I.  Anticipating
significant fading of the OA, a series of deep 300~s images was taken
with the $g$ and $R$ filters under excellent seeing conditions (FWHM
$\approx$ 0$\arcsec$.5). Inspection of the images reveals that the OA
was still bright and saturated in most of the images. Consequently,
only a single image in each of the $g$ and $R$ bands, where the OA is
not saturated, is analyzed in this study.  LRIS uses a beamsplitter to
separate the light between two arms, red and blue. Both the blue and
red cameras have a usable FOV of $\sim$ $6\farcm0\times7\farcm8$.  The
red camera used a back-illuminated Tek $2048 \times 2048$ pixel chip
with a scale of 0$\arcsec$.215 pixel$^{-1}$, while the blue camera has
a mosaic of two $2048 \times 4096$ pixel Marconi chips with a scale of
0$\arcsec$.135 pixel$^{-1}$.

Encouraged by the brightness of the OA, we also performed LRIS
spectroscopy of the OA. A preliminary analysis of the spectrum is
reported by \citet{GCN6850}, and a more detailed analysis is presented
in \S 3.5.

We performed more LRIS imaging for the OA of GRB\,071003 on 2007 October
8, 9, 10, 11, and 15, using various combinations of $u$, $g$, $V$, and
$R$ filters. The presence of the very bright star presents a
significant challenge to extracting useful data on the OA, as its
diffraction spikes change positions and intensity according to the
time and seeing conditions of the observations. Unfortunately,
observations on 2007 October 8 were adversely affected by diffraction
spikes and poor seeing, and were not usable. The data taken on 2007
October 15 are seriously affected by clouds, and do not provide an
interesting limit to the brightness of the OA, so they are not used in
this study.

We also triggered our TOO program (GS-2007B-Q-2; PI H.-W.~Chen) for
GRBs with the Gemini-S telescope and obtained $g$-, $r$-, $i$-, and
$z$-band images with the GMOS camera on 2007 Oct. 5 and 6. The GMOS
camera is equipped with three back-illuminated EEV $2048 \times 4608$
pixel chips. For our observations, the camera is used in a 2$\times$2
binning mode with a scale of 0$\arcsec$.146 pixel$^{-1}$ and a FOV of
$\sim$5$\arcmin$.5$\times$5$\arcmin$.5. Unfortunately, the 2007 October 5
images are badly affected by bleeding from the very bright star and
are not used in this study.

As part of the efforts to follow the evolution of the OA of GRB
071003, we also performed adaptive optics (AO) observations with Keck
I on 2007 October 19 \citep{GCN7010}. The details of the AO observations
can be found in \S 3.6.

\subsection{Radio Observations}

GRB\,071003 was observed with the Very Large Array (VLA)\footnote{The
NRAO is a facility of the National Science Foundation, operated under
cooperative agreement by Associated Universities, Inc.} on various
occasions. We made the observation in the B configuration array. We
used VLA source 1950+081 as phase calibrator for 4.86 GHz (C) band
observations and 2001+104 for 8.46 GHz (X) band observations.
The data were analyzed using standard data reduction routines of the
Astronomical Image Processing System (AIPS).  The first observation
took place on 2007 October 5 in the X band with flux density of $393 \pm 55$
$\mu$Jy. Since then we made six observations in the X band and three
observations in the C band (Table \ref{tab:vla}).


\section{Data Reduction} 

The bright star in the neighborhood of the OA of GRB 071003 makes it a
challenge to measure reliable photometry from the data described in \S
2. In this section we describe the methods used to minimize its
contamination.

\subsection{Photometric Calibrations}

\begin{figure}
\centerline{\includegraphics[width=3in,angle=0]{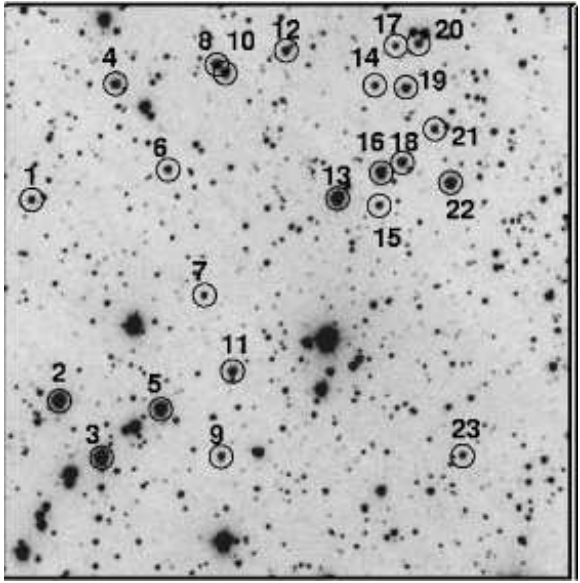}}
\caption[relation] {Finder chart for a subset of local standard
stars in the field of GRB 071003. The field of view is
6$\arcmin$.6$\times$6$\arcmin$.6.  North is up and east is to the
left. The displayed image is the unfiltered template taken with KAIT
on 2007 October 9.}
\label{fig:finder}
\end{figure}

For photometric calibrations, the field of GRB 071003 was observed in
$B$, $V$, $R$, and $I$ on two photometric nights (2007 October 7 and 8)
at Lick Observatory, using both KAIT and the Lick Nickel 1-m
telescope.  About a dozen Landolt standard-star fields (Landolt 1992)
were observed at different airmasses throughout each photometric
night. Photometric solutions to the Landolt standard stars yield a
scatter of $\sim$0.02 mag for all the filters.  The GRB 071003 field
was also observed for several sets of $BVRI$ images with different
depth on both nights. The photometric solutions are used to calibrate
a set of local standard stars in the GRB 071003 field. Because the GRB
071003 field is quite crowded, the number of calibrated local standard
stars is large (Table \ref{tab:calib}). A finder for a subset of 23
relatively bright local standard stars is in Figure \ref{fig:finder}.
As seen in Table \ref{tab:calib}, the local standard stars in the
field of GRB 071003 are well calibrated, with standard deviation of
the mean (SDOM) of $\sim$0.01 mag for all the $BVRI$ bands. We 
refer to this calibration as the ``Lick calibration" throughout the
rest of the paper.

Several Landolt standard-star fields were also observed with LRIS at
Keck I: in the $u$, $g$, and $R$ bands on 2007 October 9, and in the $V$
band on 2007 October 11.  As the number of the observed standard-star
fields is small, it is not possible to derive a complete photometric
solution for either night. Since the GRB field was observed at similar
airmasses with some of the standard-star fields, we can treat the LRIS
filters as standard and derive the magnitudes for the local standard
stars via differential photometry. Unfortunately, this procedure
suggests that the 2007 Oct. 9 night was not photometric, as different
standard-star observations yield somewhat different zero points. The
2007 October 11 night was photometric, but only the $V$-band standard stars
were observed.

We elected to use the Lick calibration as the foundation for all the
photometric calibrations, except in the case of the $u$ band. The
Lick-calibrated magnitudes are in $BVRI$, and can be reliably
converted to the $g$, $r$, and $i$ bands using color transformation
equations \citep{Jester+2005}.  The conversion to the $z$ band
\citep{Rodgers+2006} is somewhat problematic, and as a result we adopt
a relatively large uncertainty for the converted magnitudes. For the
$u$ band, only two standard-star fields were observed with LRIS on
2007 Oct. 9, and they give a difference of 0.30 mag in the zero
points. We chose to calibrate the GRB 071003 field with the
standard-star field that is closer in time of GRB observation, but
we added an uncertainty of 0.30 mag to all the calibrated magnitudes. We
note that the true error for the $u$-band calibration may be higher than
0.30 mag due to the nonphotometric conditions on 2007 October 9.

\subsection{KAIT Data Reduction}

The KAIT data were automatically processed with bias and dark current
subtraction and flat-fielding. The PSF of the OA is seriously
affected by the bright star which is less than 10 pixels away in the
KAIT images. Consequently, normal PSF-fitting photometry cannot fit the
peak and background of the OA simultaneously to produce a reliable
measurement.

\begin{figure}
\centerline{\includegraphics[width=3in,angle=0]{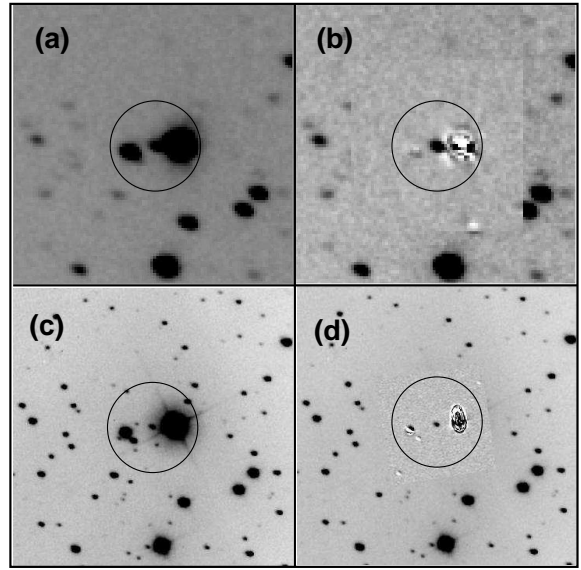}}
\caption[relation] {Illustration of using image subtraction to remove
the contamination of the bright nearby star to the OA of GRB 071003.
The KAIT image subtraction code is demonstrated here. (\textit{a}) An
$80\arcsec \times 80\arcsec$ section of the original 20~s unfiltered
KAIT image of the OA taken at $t = 431$~s; (\textit{b}) the same section after
image subtraction of the central $50\arcsec \times 50\arcsec$ using an
unfiltered template image after the OA has faded; (\textit{c}) an $80\arcsec
\times 80\arcsec$ section of the combined unfiltered AEOS image at $t
= 5002.6$~s; and (\textit{d}) the same section after image subtraction of the
central $30\arcsec \times 30\arcsec$ using a hand-made template
image. See text for more details. }
\label{fig:imagesub}
\end{figure}

We use image subtraction to remove the contamination of the bright
star.  To generate template images for subtraction, KAIT imaged the 
GRB 071003 field in the unfiltered mode and in the $V$ and $I$ filters 
for the next several nights after the burst.  To make sure the bright 
star is not saturated, short (5~s) exposures were used, 
and 50--100 images for each filter were acquired to ensure high S/N 
in the combined images. As discussed in
\S 4, the GRB OA was still reasonably bright in the second night after
the burst, so we used the images obtained at 4--6~days after the burst
as the template for the field without significant OA contribution.
Our image subtraction code is based on the ISIS package
\citep{Alard+1998} as modified by B.\ Schmidt for the High-$z$ Supernova
Search Team \citep{schmidt98}.  An illustration of the image
subtraction is presented in the top panels of Figure
\ref{fig:imagesub}.

The Lick calibration was used to transform the KAIT instrumental
magnitudes to the standard Johnson $V$ and Cousins $I$ passbands, with
proper color terms measured from the photometric nights. We also find
that the combination of the KAIT optics and the quantum efficiency of
the FLI CCD camera makes the KAIT unfiltered observations mostly mimic
the $R$ band. During the two photometric nights, unfiltered
observations of the Landolt standard-star fields were also performed.
Analysis of these images indicates that the KAIT unfiltered magnitudes
can be effectively transformed to the $R$ band, with a relatively
large color term and an rms of $\sim$0.05 mag,
similar to the earlier results we reported
\citep{Li+2003a,Li+2003b,Butler+2006}.

\begin{figure}
\centerline{\includegraphics[width=3.5in,angle=0]{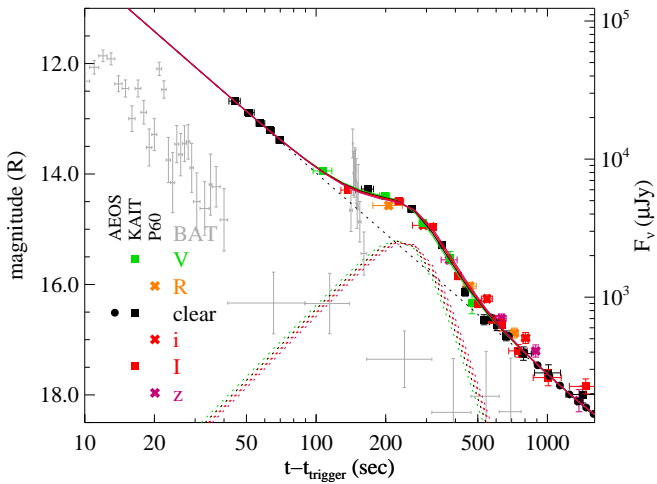}}
\caption[relation] {Early-time light curve of the optical afterglow of
GRB~071003 using KAIT photometry, supplemented by observations from
P60 and AEOS.  The gamma-ray light curve from the BAT is overplotted
in gray (scaled arbitrarily).  A clearly additive ``bump'' at
100--500~s is apparent.  Photometric follow-up observations continued after 
2000~s with P60 and AEOS, as well as with Gemini and Keck in subsequent
nights; the complete 16-day optical light curve is presented in Figure
\ref{fig:lc}.}
\label{fig:earlylc}
\end{figure}

To increase the S/N, the late-time KAIT images of GRB 071003 were
combined into groups of three to eight images. The final KAIT photometry for
the GRB 071003 OA is listed in Table \ref{tab:kait}. The reported
error bars are the uncertainties in PSF-fitting photometry and those
in the calibration process, added in quadrature.  A plot of the KAIT
photometry, along with measurements from other telescopes during the 
same timespan (with BAT data overplotted and fitted by a chromatic model
described in \S~4.3) is presented in Figure \ref{fig:earlylc}.

\subsection{AEOS Data Reduction}

The ABC images were processed using dark subtraction only.  Because of
highly variable stray light and vignetting, we did not apply a flat
field to these images.  We used SExtractor \citep{Bertin+1996} to find
all the sources in the images, from which we were able to determine
the astrometry.

We employed the NN2 flux difference method (\citealt{Barris+2005};
hereafter the NN2 method) for constructing the AEOS light curve. The
NN2 method also uses image subtraction to measure the fluxes for a
variable source, but it does not designate one particular image as the
template. Instead, given $N$ total observations, the NN2 method solves
for the vector of fluxes from the individual images using the
antisymmetric matrix of flux differences from the $N(N-1)/2$ distinct
possible subtractions. Compared to the template image subtraction
method, the NN2 method takes all the available information from the
images into account, and is less susceptible to possible noise
associated with a single template image.  To avoid a large number of
image subtractions, we combined the original 238 AEOS observations
into 39 images. For the first 228 images, each set of six consecutive
images is combined into one.  The last 10 images are combined into a
single image.  We compared the results from the NN2 method to those
from a traditional template image subtraction method (bottom panels of
Figure \ref{fig:imagesub}) and found them to be consistent with each
other.

To calibrate the AEOS data to the standard photometry system, we used
the KAIT $R$-band data during the overlap period and assume that the
unfiltered AEOS data have no color term to the $R$ band.\footnote{We
attempted to quantify the color term of the unfiltered AEOS data to
the standard $R$ system using the local standard stars in the field of
GRB 071003, but found no apparent correlation between the scatter of
the (unfiltered $-$ $R$) differences versus the colors of the stars.}
The final AEOS photometry is listed in Table \ref{tab:aeos}. The
reported error bars are only those output by the NN2 method, and do
not include a possible large systematic error due to calibration. If
the throughput of the AEOS telescope in the unfiltered mode is not
drastically different from that of KAIT, we estimate the systematic
error to be $\sim$0.07 mag when the GRB OA was bright ($t < 20$ minutes),
and $\sim$0.15 mag when the GRB became faint ($t > 40$ minutes). The
systematic errors can be much higher if the unfiltered throughput is
very different for the two telescopes.

\subsection{Keck I/Gemini-S Data Reduction}

Due to the large aperture of the Keck I and Gemini-S telescopes, the
bright star close to the GRB 071003 OA produces numerous diffraction
spikes, as well as two large blooming spikes along the readout
direction.  Because the orientation, width, and intensity of the
spikes change with the seeing conditions, the exposure duration, and
the time of the observations, it is difficult to cleanly remove them
using the template image subtraction or the NN2 method. However, due
to the high resolution of these images, the spikes are well sampled
and show distinct axial symmetry.  We developed a saturation spike
subtraction method, in which we divide the image of the bright star in
half, flip the right side, and subtract it from the left side. Due to
the symmetry in the spikes, this subtraction process leaves a
reasonably clean region around the GRB OA.  PSF-fitting photometry was
then performed on the GRB OA in the spike-subtracted images, and on a
series of local standard stars. The Lick calibration is used to
calibrate the Keck I and Gemini-S instrumental magnitudes to the
standard system.

The final Keck I and Gemini-S photometry is reported in Table
\ref{tab:keck}.  The error bars of the magnitudes are the
uncertainties from the PSF-fitting photometry and those in the
calibration process added in quadrature.  One special data point is
the Keck I HIRES guider image at $t = 9523.7$~s because it bridges the
early KAIT/AEOS data to the late-time Keck I and Gemini-S
observations. The GRB OA was well detected in the guider image, but
because the image has a small FOV and is unfiltered, photometric
calibration becomes particularly difficult.

We have used three methods to calibrate the measured instrumental 
magnitude of the OA after the guider images were processed with the
saturation spike subtraction method: differential photometry between
the AEOS unfiltered data and the guider images, photometric
calibration to about half a dozen stars in the HIRES guider images
using the KAIT unfiltered images, and photometric calibration to these
stars using the Keck I $R$-band images. The measured $R$-band
magnitudes from these three methods show a scatter of $\sim$0.25 mag,
and their average value and uncertainty are listed in Table
\ref{tab:keck}.

\subsection{P60 Data Reduction}

The P60 data reduction is presented in this section because it employs
several methods (illustrated in Figure \ref{fig:imagesub}) discussed
earlier in the paper. We obtained template images for the field after
the OA of GRB 071003 has faded. However, the saturation spikes of the
bright star close to the GRB ruined the template images in the $R$ and
$i^\prime$ bands, so we were only able to run image subtraction for
the data in the $g$ and $z^\prime$ bands. We also employed the
saturation spike subtraction methods as described in \S 3.4. Although
P60 does not have the resolution of the Keck I and Gemini-S telescopes,
subtraction of half of the saturation spikes helped to clean up the
background of the OA considerably.

We also applied a third method to reduce the P60 data. Due to the
richness of stars in the GRB 071003 field and the large field of view
of the P60 camera ($12\arcmin.9\times12\arcmin.9$), we were able to
pick a star that is close in brightness (within 0.1 mag in all
filters) and thus has similar saturation spikes to the bright star
close to GRB 071003. The chosen star is located at $\alpha$ =
$20^h07^m14^s.84$, $\delta$= $+10^\circ53\arcmin59\arcsec.8$ (equinox
J2000.0), which is $136\arcsec.7$ west and $172\arcsec.0$ south of the
GRB 071003 OA.  By slightly scaling the PSF of this bright star and
subtracting it from the star close to the GRB, we were able to largely
remove the complicated background around the GRB OA.

PSF-fitting photometry is applied to the images after different ways
of image subtraction, and the Lick calibration is used to calibrate
the instrumental magnitudes into the standard system. The final
photometry from the P60 data is listed in Table \ref{tab:p60}, which
is the average of the spike and bright star subtraction methods.  The
results from the template image subtraction method are not considered
because the method can only be applied to a subset of filters, but
they are consistent with the other two methods within measurement
uncertainties.

\subsection{Keck AO Data Reduction}

\begin{figure}
\centerline{\includegraphics[width=3in,angle=0]{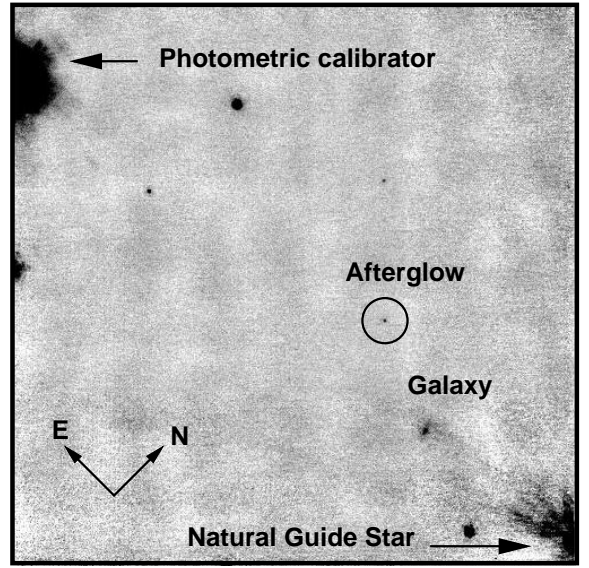}}
\caption[relation] {NGS AO image
of the GRB~071003 field taken with Keck II on 2007 October 19, 16~days after
the burst. The FOV is approximately $12\arcsec \times 10\arcsec$. The
afterglow is well detected with $K' = 21.58 \pm 0.03$ mag. No 
host-galaxy emission is detected.  }
\label{fig:keckAO}
\end{figure}

On 2007 October 19 (starting at UT 05:14) we observed the GRB 071003 OA
with the NIRC2 \citep{VanDam+2004} narrow-field camera (0$\arcsec$.01
pixel$^{-1}$) on Keck II using natural guide star adaptive optics (NGS
AO).  While the extremely bright nearby star greatly complicated the
optical analysis, it was ideal to be used as the natural guide star
during NGS AO imaging.  We took 15 science exposures, each of 60~s and
2 coadds, resulting in a total integration time of 30 minutes.  The images
were reduced using standard techniques, including dark subtracting,
flat fielding, and filtering for deviant pixels.  Each frame was
dewarped using the recommended method for NIRC2, and the resulting
images were registered to a common origin and combined.

The GRB OA is well detected 2 weeks after the burst, as shown in the
final combined image in Figure \ref{fig:keckAO}.  To measure the
brightness of the OA, we created a model of the PSF using short-exposure, 
unsaturated images of a nearby Two Micron All Sky Survey (2MASS) star ($K_s = 12.011
\pm 0.024$ mag, $d = 7.8\arcsec$), taken immediately prior to the
science exposures.  We then subtracted this model PSF from the OA.  
With the same 2MASS star as the photometric 
calibrator, we measure the OA to have $K^\prime$ = 21.65 $\pm$ 0.10 Vega mag.
(Galactic reddening of $A_{K^\prime} \approx 0.05$ mag is negligible
along this sightline and has not been applied.)

\subsection{Keck LRIS Spectroscopy Reduction}

We obtained low-resolution optical spectroscopy of the optical
afterglow of GRB 071003 on 2007 October 4.335 using the LRIS on the
Keck I telescope.  A pair of 600~s dithered exposures was taken
under clear conditions at airmass 1.2 with 0.6\arcsec\ seeing.  We
used both the blue and red arms of LRIS, with the light split by the
D680 dichroic.  The 300/5000 grism on the blue side gave a spectral
resolution of 8.4~\AA\ over the range 3300--6500~\AA .  We used the
600/10000 grating to achieve 4.1 \AA\ resolution over the range
6500--8630~\AA .  The spectrophotometric standard star Feige 110
\citep{Stone1977} was observed the following night in the same setup.
Intermittent clouds were present the night of the standard-star
observation, so the absolute flux scale is unreliable.  

The long, 1.0\arcsec-wide slit was oriented at a position angle of
10$^\circ$ for the afterglow observations, which was not the
parallactic angle \citep{filippenko82}.  However, the Cassegrain
Atmospheric Dispersion Compensator module \citep{Phillips+2006} was
mounted, so the derived spectral shape should be reliable.  The
exception is in the spectral range of 6000--6500~\AA, where
second-order blue light contamination is prominent in the spectrum of
the standard star.  An attempt was made to correct for the
contamination, but the spectral slope in this section is more
uncertain than in the rest of the spectrum.  We also fitted an
extinction-corrected power law to the flux-calibrated spectrum
(excluding line and second-order contaminated regions) in an attempt
to estimate the spectral slope, but the estimated slope of $f_{\nu}
\propto \nu^{-0.87}$ differs significantly from the spectral slope
estimated from multi-band late-time photometry (\S 4.7).  This may be
due to continuum contamination from the nearby star in the spectrum
(which is difficult to properly remove), so we do not further consider
this spectroscopic spectral index.

\begin{figure}
\centerline{\includegraphics[width=3.5in,angle=0]{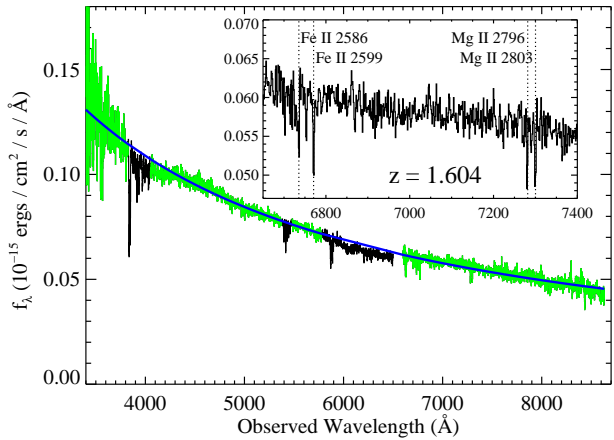}}
\caption[relation] {Spectrum of the GRB 071003 afterglow covering the
full observed spectral range.  The spectrum has been flux-calibrated
and corrected for Galactic reddening of $E(B-V)=0.148$ mag.  The inset
shows an expanded view of the region surrounding the Fe and Mg
absorption system at the burst redshift.  A power-law continuum was
fitted to the regions of the spectrum shown in green, chosen to avoid
strong absorption lines and the wavelength range contaminated by
second-order blue light.  The thick solid blue line shows the
resultant fit ($f_{\lambda} \propto \lambda^{-1.13}$, or $f_{\nu}
\propto \nu^{-0.87}$), but it differs in slope from our more reliable 
fit to the broadband photometry; thus, it is used only to normalize
the spectrum.}
\label{fig:kecksp}
\end{figure}

The largely featureless spectrum (Figure~\ref{fig:kecksp}) has a
S/N $>5$ ${\rm pixel}^{-1}$ down to $\sim$3500~\AA.  There is no apparent
absorption by the intergalactic medium at these wavelengths,
yielding an upper limit to the redshift of the burst of $z_{GRB} <
(3500/1216) - 1 = 1.88$.  Numerous metal-line absorption lines (but no
emission lines) are visible in the spectrum.  We have fitted the
equivalent widths of all $\gtrsim 5 \sigma$ features in the normalized
spectrum using a Gaussian profile and report the rest-frame values in
Table~\ref{tab:lines}.

\begin{figure}
\centerline{\includegraphics[width=2.7in,angle=90]{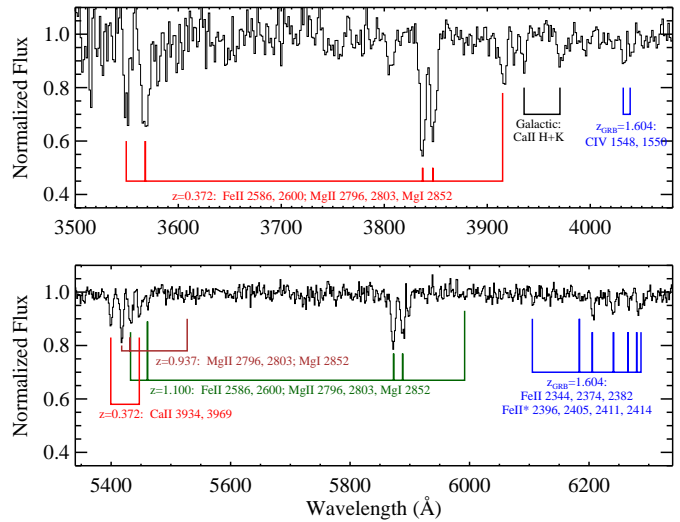}}
\caption[relation] {Portions of the normalized Keck LRIS spectrum of
the GRB~071003 afterglow.  We mark the positions of several metal
absorption-line features from four distinct extragalactic systems
including a series of \ion{Fe}{2} and \ion{Fe}{2}* transitions
associated with the host galaxy of GRB~071003 ($z_{GRB} = 1.604$).
Note that the \ion{Ca}{2} doublet marked as Galactic may be due to the
very bright Galactic star offset by 6.5$''$ from GRB~071003 as opposed
to the Galactic ISM.}
\label{fig:intervene}
\end{figure}

\begin{figure}
\centerline{\includegraphics[width=3.5in,angle=0]{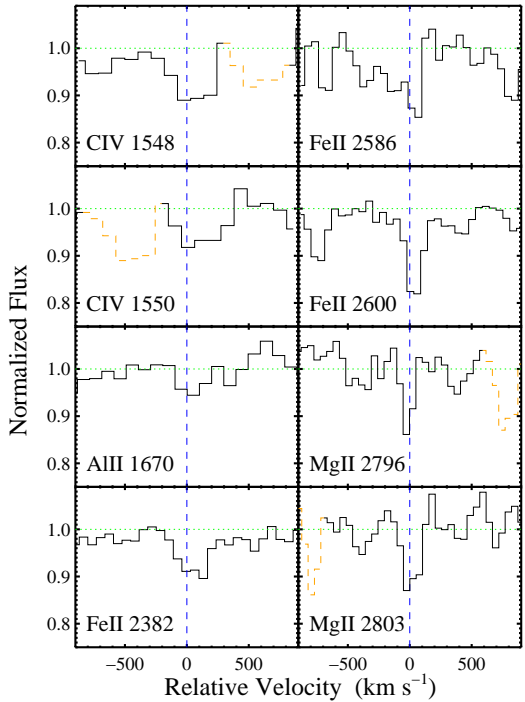}}
\caption[relation] {Velocity plot of strong, resonance-line
transitions for gas associated with GRB~071003 ($z_{GRB}=1.60435$).
These lines are very weak (note the ordinate scale) with rest-frame
equivalent widths of 100--200~m\AA\ (Table~\ref{tab:lines}).
Indeed, the \ion{C}{4} absorption is the weakest yet reported for a
GRB afterglow \citep{pcw+08}.}
\label{fig:grbvel}
\end{figure}

We previously presented \citep{GCN6850} analysis of this spectrum,
identifying \ion{Mg}{2} absorption systems at $z=0.372$ and
$z=1.100$.  A VLT spectrum acquired the same night \citep{GCN6851}
identified a third absorption system at $z=0.937$, which is confirmed
by our observations.  These are the only strong absorption systems in
the data, and previously we considered it likely that the $z=1.100$
system originated from the host galaxy (Figure \ref{fig:intervene}).
Surprisingly, however, a more thorough investigation revealed a
fourth, weak absorption system at a higher redshift of $z=1.604$
(Figure \ref{fig:grbvel}).  Contrary to our expectation, the gas at
this redshift has the weakest \ion{Mg}{2} absorption of the four
systems.

This is remarkable: absorption lines associated with GRB environments
are generally very strong with rest-frame equivalent widths exceeding
several angstroms \citep{sff03,pcw+08}.  Figure~\ref{fig:intervene}
also indicates, however, the presence of fine-structure \ion{Fe}{2}
transitions at this redshift.  With the exception of active galactic
nucleus environments, these transitions have only been identified in
gas surrounding the GRB phenomenon \citep{pcb06}.  These transitions
are excited by the GRB afterglow itself through indirect ultraviolet
pumping \citep{pcb06,vls+07} of gas in the interstellar medium (ISM)
of the host galaxy.  Altogether, the coincidence of (1) the absence of
any higher-redshift absorption systems in our spectrum, (2) the
positive detection of fine-structure \ion{Fe}{2} transitions, and
(3) the absence of intergalactic medium absorption at $\lambda >
3500$~\AA\ establishes $z=1.604$ as the redshift of GRB~071003.

It might seem unusual to have detected fine-structure \ion{Fe}{2}
transitions in such a late-time spectrum ($t \approx 24.3$~hr).
Because the lines are excited by the GRB afterglow, they will decay as
the afterglow fades on hour-long timescales
\citep{dcp+06,vls+07,DElia+2008}.  The presence of fine-structure
transitions in our spectrum, however, is consistent with the late-time
rebrightening of GRB~071003 provided that the gas lies within a few kiloparsecs of
the GRB.  In Figure~\ref{fig:grbvel} we present a velocity plot of
strong resonance-line transitions for $z = z_{GRB}$.  We report the
positive detections of \ion{C}{4}~$\lambda$1548,
\ion{Fe}{2}~$\lambda\lambda$2382, 2586, 2600, and
\ion{Mg}{2}~$\lambda$2803, and we note probable but statistically
insignificant absorption at \ion{Al}{2}~$\lambda$1670 and
\ion{Mg}{2}~$\lambda$2796.  The rest-frame equivalent widths are
among the lowest ever recorded for the ISM surrounding long-duration
GRBs.  The equivalent width of \ion{Mg}{2}, for example, is fully an
order of magnitude below the general population \citep{Cenko+2008},
with the sole exception of GRB~070125, and
the equivalent width for the \ion{C}{4} gas ($W_{1548} = 0.22 \pm
0.06$~\AA), represents the lowest measurement to date \citep{pcw+08}.

\section{Results and Modeling}

\subsection{Light Curve: General Observations}

\begin{figure*}
\centerline{\includegraphics[width=6in,angle=0]{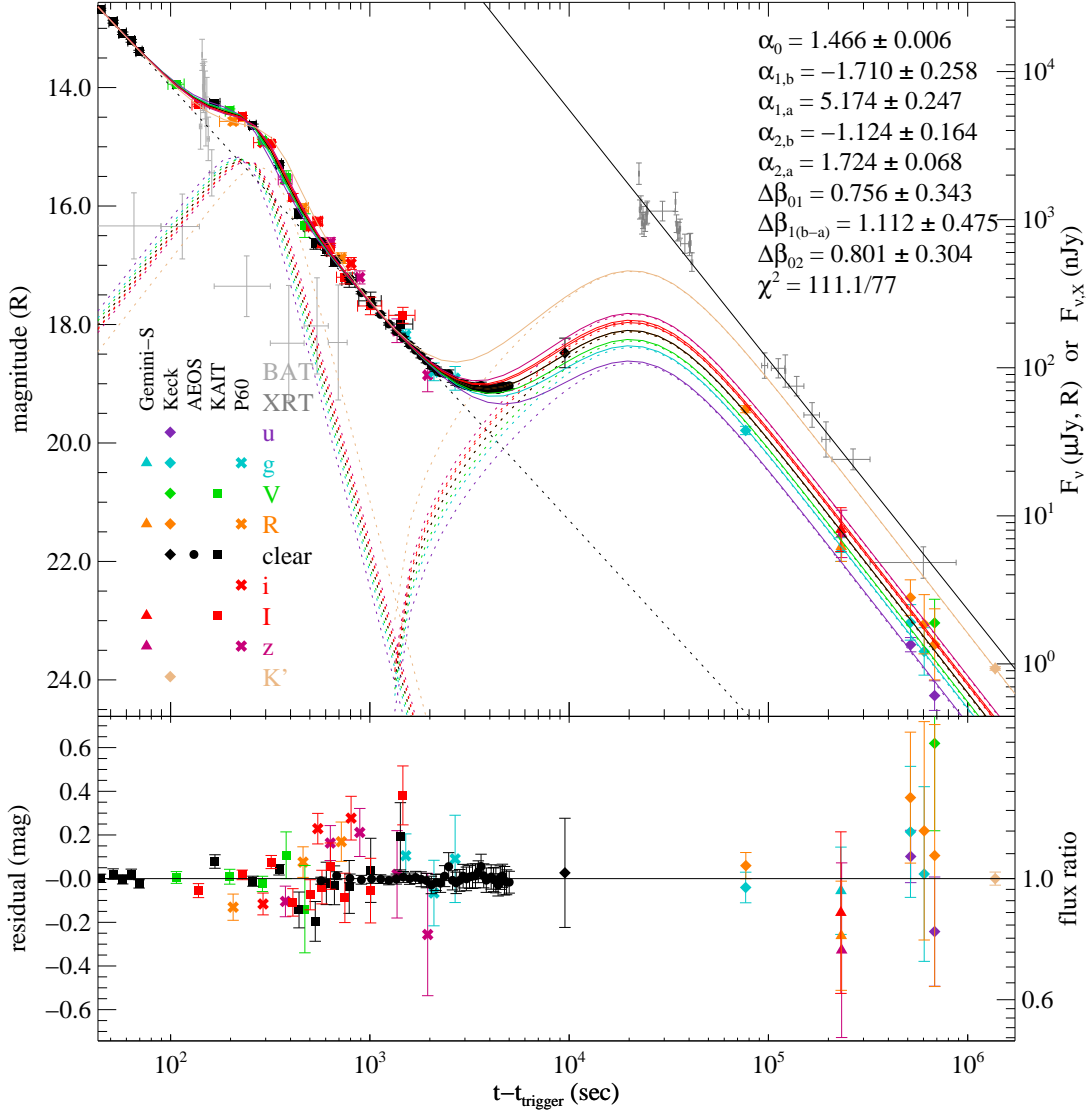}}
\caption[relation] {Multi-color, early through late-time light curves
of the OA of GRB~071003.  The magnitudes are offset
according to their early-time colors, showing the color evolution
between early and late times.  Overplotted colored curves indicate the
best-fit three-component, color-evolution model described in the text;
the dashed lines represent the individual components that compose this
model (a uniform power-law decay, a chromatic early-time bump, and a
monochromatic late-time rebrightening).  The X-ray and gamma-ray
afterglows are also overplotted for comparison.  The gamma-ray light
curve is scaled arbitrarily; if scaled based on the likely
gamma-to-X-ray spectral index it would fall on or near the
extrapolation of the X-ray light curve back to early times.}
\label{fig:lc}
\end{figure*}

The multi-color photometric evolution of the GRB 071003 OA is shown in
Figure \ref{fig:lc}, fitted by our preferred model (described later).
Visual inspection of the light curves reveals what appear to be three
distinct components: an overall power-law decline that has already set
in by the very first measurement at 42~s, a small ``bump'' feature at
$\sim$120--600~s, and then a dramatic, but unfortunately not well
sampled, rebrightening starting around 3000~s that dominates the
remainder of the evolution.

The bump feature appears to be additive only: fitting a single power
law to measure the decay index ($t^{-\alpha}$) for the clear-band data
both before this period and after it, the power-law indices ($\alpha =
1.47$ and $\alpha = 1.49$, respectively) are fully consistent with
each other and with the overall decay index over both periods
($\alpha = 1.48$).

The rebrightening is more difficult to characterize.  We have no
observations between the Keck I HIRES guider point at $t \approx 2.6$
hr and our observations the second night; moreover, the points
reported in the GCNs are highly discrepant.  An optical $R$-band limit
is reported at $t \approx 4$~hr by \citet{GCN6846}, which seems to
contradict the rebrightening trend suggested by the AEOS data and
guider point. It is unlikely that the OA would show such a dramatic drop
($> 3$ mag) in a short time interval at such late times, so we suspect that
the OA might be heavily contaminated by the bright nearby star and was
not resolved in the Lulin 1-m telescope images of \citet{GCN6846}.  On
the other hand, the $U$-band detection at $t \approx 7.5$ hr reported
by \cite{GCN6840} supports a rebrightening but is several magnitudes
above the extrapolated light curve at this time, seemingly far too
bright to be consistent with our observations.  Calibration and the
contamination from the bright star are the likely causes of the
discrepancy.

\subsection{Optical to Gamma-Ray and X-Ray Comparison}

The BAT and XRT light curves we derive for GRB~071003 are also shown
in Figure \ref{fig:lc}.  Unfortunately, because \textit{Swift} was
still in the process of returning to normal operations after its gyro
failure \citep{GCN6760}, automatic slewing to GRB~071003 was disabled
at the time when the GRB was detected. As a result, there were no
prompt XRT observations for GRB 071003, leaving a long gap in the
gamma-ray/X-ray light curve at $t = 200$--20000~s.  In
particular, there are no X-ray observations until approximately the
peak of the rebrightening in the optical band.  Nevertheless, direct
comparison of the data available reveals three relevant facts.

First, there is no obvious optical prompt counterpart to the last
spike of the gamma-ray light curve.  However, this spike is nearly
contemporaneous with the much more slowly rising optical bump feature;
we return to this possible connection in our later modeling (\S
4.5).

Second, at late times the X-ray light curve declines as a power law
with decay index consistent with that observed in the optical.  A
simple power law fits the data well, with a best-fit decay index of
$\alpha_X = 1.68 \pm 0.04$. In addition, the late-time OA behavior
(after $t \approx 5 \times 10^4$~s) is consistent with a single
power-law decay with an index of $\alpha_O = 1.72 \pm 0.07$, fully
consistent with this value.  As we note later, an extrapolation
of the X-ray spectral index is also consistent with the optical
observations, suggesting that at late times there is no need for an
additional X-ray contribution (such as inverse Compton) or large
amounts of host-galaxy extinction.

Finally, while the gamma rays are scaled arbitrarily in Figure
\ref{fig:lc}, we note that if we extrapolate the gamma-ray spectrum
into the X-rays to compare the BAT and XRT light curves, the evolution
between the end of the prompt emission and the start of the XRT
observations is nearly consistent with a simple extension of the
late-time XRT power law back to earlier times, without a need for a
rebrightening or break.  However, \textit{Swift} has shown previously
\citep{Nousek+2006} that early-time X-ray light curves can conceal a
wide variety of complex features, so we will not speculate further as
to whether or not this was actually the case.

\subsection{Detailed Optical Modeling}

The procedure used to model the optical light curve is generally the
same as that employed by \citet{Perley+2008}, but further generalized.
For our fit model, we employ an unbroken power-law decay (component 0)
plus two \citet{Beuermann+1999} functions (broken power-law pairs,
components 1 and 2), but allow for different values of the functional
parameters for each filter and component.  The functional form is

\begin{equation}
\begin{array}{rcl}
F_\nu & = & F_{0,\nu}\,(t- dt_0)^{-\alpha_0} \\
         && + F_{1,\nu}\,(0.5 \,(\frac{t - dt_1}{t_{p1}})^{-s_1 \alpha_{1,b}} + 0.5 \,(\frac{t - dt_1}{t_{p1}})^{-s_1 \alpha_{1,a}})^{-\frac{1}{s_1}} \\ 
         && + F_{2,\nu}\,(0.5 \,(\frac{t - dt_2}{t_{p2}})^{-s_2 \alpha_{2,b}} + 0.5 \,(\frac{t -  dt_2}{t_{p2}})^{-s_2 \alpha_{2,a}})^{-\frac{1}{s_2}},
 \end{array}
\end{equation}

\noindent 
where for component 0, $\alpha_0$ is the power-law decay
index, and $dt_0$ is an adjustment to the \textit{Swift}/BAT trigger
time. For cmponent 1, $\alpha_{1,b}$ and $\alpha_{1,a}$ are the
power-law decay indices for the rising and declining components,
respectively, $dt_1$ is an adjustment to the \textit{Swift}/BAT
trigger time, $t_{p1}$ is the time of the peak flux, and $s_1$ is the
sharpness parameter. Component 2 has a similar function as component
1.

Fitting this function with no constraints generates unrealistic
results because of non-uniform sampling in different filters.
However, we can make the following physically motivated assumptions to
tie specific parameters and produce more physically meaningful
results.

\begin{enumerate}
\item{We assume that the temporal decay index at any given time is
independent of the filter, as is implicit in the notation ($\alpha$
does not depend on $\nu$).  This means that the color of a component
cannot change except while the light curve of that component is
breaking.}

\item{Most importantly, we assume that differences between the spectra
of the various model components can be described by changes in the
power-law index of the intrinsic spectrum, modified by an arbitrary,
but fixed, extinction law.  Mathematically, this constraint is
expressed as $F_{i,\nu} = \nu^{\Delta\beta_{ij}} F_{j,\nu}$.
Physically, this assumption requires that external effects such as
extinction, which might cause the spectrum of any component to deviate
from a power law, affect all components equally and are not
time-dependent.  The extinction law itself (as well as the
\emph{absolute} underlying index of any specific component) is fully
general and can be fitted according to various models later.}

\item{In addition, we assume that the rising segments of each
component are also power laws, but not necessarily the same power laws
as the falling segment, to allow for chromatic breaks.  This imposes
the following condition: $(\frac{t_{p},x}{t_{p},y}) =
(\frac{\nu_y}{\nu_x})^{\Delta\beta_{ba}/\Delta\alpha_{ba}}$.  Here $b$
and $a$ refer to ``before'' and ``after'' the break of a specific
component (0, 1, or 2), where the component index is omitted for
clarity, and $x$ and $y$ refer to two different filters.}
\end{enumerate}

Fitting is performed, under these assumptions, using the IDL package
mpfit \footnote{\url
http://cow.physics.wisc.edu/$\sim$craigm/idl/idl.html .}.

The assumptions involved in these constraints are, of course,
oversimplifications for the full array of models that might be
considered.  In particular, this model allows only one break per
component, but with an evolving synchrotron light curve plus a jet we
may expect as many as three.  However, it has the advantage of being
simple and generates a single physically motivated parameter
quantifying color change over each component.

We perform a variety of fits under varying combinations of
assumptions.  Some of the possibilities we considered include
the following:

\begin{enumerate}
\item{Forcing the bump (component 1) to have the same color as the
uniform decay (component 0), or allowing it to be a different color
overall.}

\item{Forcing the bump itself to be achromatic over its evolution, or
allowing it to contain a chromatic break.}

\item{Forcing the late rebrightening (component 2) to have the same
color as the uniform decay, or allowing it to have a different color.}

\item{Fixing $dt_0$ for the early steep decay to be zero (the BAT
trigger time), or allowing it to be free to vary.}

\item{Fixing $dt_1$ for the bump component to be zero, to be equal to
the beginning of the prompt-emission pulse that is nearly
contemporaneous with it, or allowing it to be free to vary.}

\item{Fixing $dt_2$ for the late rebrightening to be zero, or allowing
it to be free to vary.}
\end{enumerate}

The results under various combinations of these assumptions are
presented in Tables \ref{tab:colorchange} and \ref{tab:t0}.  We
discuss the implications of these results in the remainder of the
paper.

\subsection{Color Change} 

Detection of a GRB afterglow in filtered observations during prompt
emission, as was the case here, is rare.  The situation is even more
intriguing since our multi-color prompt OA observations show an
apparent bump feature (component 1) that is nearly contemporaneous
with a rebrightening pulse in the gamma-ray light curve.  Therefore,
it is of great interest to attempt to measure the color of Component
1.  By the same token, we have good spectral coverage of the afterglow
both during the primary normal decay and during the fading of the
dramatic late rebrightening, and any color difference may shed light
on the origin of these features.

We tested for color differences in three places: between component 0
(rapid decay) and component 1 (bump), between component 0 and
component 2 (rebrightening), and over the break of component 1 itself
(since the rising spectral index may differ from the falling spectral
index).  In all cases we find evidence for color variation, although in
each case only at the $\sim2\sigma$ level.  The fading component of
the bump is redder than the fading component of the uniform decay by
$\Delta\beta$ = 0.75 $\pm$ 0.34, the bump feature is chromatic with a
shift from the rising to falling component of $\Delta\beta$ = 1.11
$\pm$ 0.47, and the rebrightening (for which we only have color
information during the fading component) is also redder, by
$\Delta\beta$ = 0.84 $\pm$ 0.31.

One must be somewhat cautious in interpreting these results --- since
different filters sample the data differently, systematic errors that
affect only one portion of the light curve can masquerade as color
change. Data reduction for GRB 071003 was also challenging due to the
presence of the nearby bright star, as detailed in \S 3. In addition,
we note that the degree of spectral index shifts noted is dependent
on the model.  In spite of these considerations, however, we feel that
our conclusion of color change is reasonably secure in each case.

\subsection{Energy Injection Times}

It is often unclear what time is most appropriate to use as $t_0$ when
fitting a power law to a GRB afterglow.  Thanks to the extremely
early-time clear-band data, it is possible to fit $t_0$ and constrain
this within a few seconds in the case of GRB 071003.  This fit,
notably, gives a $t_0$ of exactly the trigger time ($dt_0 =
-0.01 \pm 3.01$~s).  The gamma-ray light curve (Figure
\ref{fig:grblc}) fluence is strongly dominated by the initial pulse,
which rises sharply and peaks within a few seconds, so this is not
necessarily surprising.

Some authors \citep{Blake+2005, Vestrand+2005, Vestrand+2006,
Yost+2007} have presented evidence of an optical component rising
coincident with the prompt emission, although significantly longer
lasting.  We can analyze whether the bump component observed in 
GRB~071003 may be such a
feature by determining whether or not it can be fitted with a pulse that
rises abruptly, contemporaneous with the prompt emission.  While our
power-law model is somewhat simplified and the sampling of the rise is
extremely poor, we find that it generally does not: the best-fit $t_0$
is intermediate between the trigger time and the time of the prompt
emission spike ($\sim$125~s) at $dt_1 = 60 \pm 20$~s.  This is a
model-independent result, although it rests mostly on one data-point:
the initial $V$-band measurement, representing an integration from
97 to 117~s after the BAT trigger ($\sim$18~s before the rise of the
prompt emission spike), lies 0.14 mag above a simple power-law
extrapolation from regions of the data excluding the bump, compared to
a photometric error of only 0.03 mag.  While it is possible to
envision scenarios where a relatively slow optical rise might follow a
gamma-ray pulse (any broadband feature with hard-to-soft evolution, or
perhaps a late internal shock that later collides with and energizes
the external shock), no model to our knowledge can explain why an
optical flare would precede a gamma-ray pulse, so we take this as
evidence that the two features are physically unconnected.

While our sampling around the rise and peak of the late-time rebrightening is
poor (and dominated by the difficult-to-calibrate AEOS and HIRES guider images), 
we can also attempt to fit the $t_0$ for the rebrightening
component.  This is significantly different from $t = 0$, with a 
best-fit initial time of $dt_2 = 1245 \pm 311$~s.  (This is well short
of its peak time of approximately 20~ks, so the effect on the light
curve is minor.)  No prompt-like fluctuations or other features are
observed in the light curve in this region.

\subsection{Radio Modeling}

\begin{figure} 
\centerline{\includegraphics[width=3.5in,angle=0]{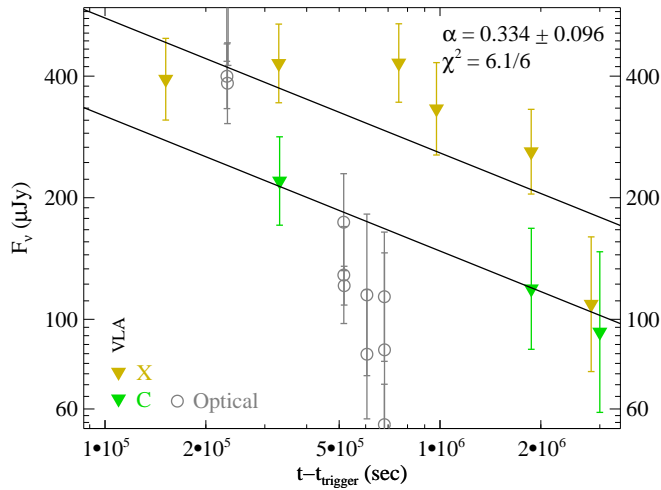}}
\caption[relation] {\small VLA radio light curve fitted to an unbroken
power law.  The uncertainties in the measurements have been increased
compared to their statistical values to take into account the effect
of interstellar scintillation.  Some contemporaneous late-time optical
points (scaled arbitrarily) are shown for comparison.}
\label{fig:vlalc}
\end{figure}

GRB 071003 is rare among \textit{Swift} bursts for having a bright
radio afterglow.  We were able to successfully detect the afterglow at
two frequencies and several epochs spanning $\sim$2--20~days after the
burst, including observations nearly contemporaneous with our optical
data.  The data are plotted in Figure \ref{fig:vlalc}.

This GRB is not far off the Galactic plane, and the radio observations
are affected by scintillation.  Following \citet{Walker1998,
Walker2001}, the afterglow is in the strong scattering regime for both
X and C bands.  An approximate modulation index (which estimates the
fractional rms variation) is 0.4 in the C band and 0.6 in the X band,
over a refractive timescale of $\sim$0.5~days in the X band and 2~days in
the C band.  This is longer than any integration (so the error is not
reduced by integration time) but shorter than the interval between
exposures (so errors are uncorrelated).

Radio data were fitted using both an unbroken power-law model and a singly
broken power-law model.  We attempted the fit both before including
uncertainties due to scintillation and with an additional 40\% flux
error added to all C-band points and 60\% error added to all X-band
points.

Without the additional flux errors, the unbroken model is a poor fit,
with $\chi^2/\nu = 15.32/6$.  A single, monochromatic break
improves the fit dramatically ($\chi^2/\nu = 2.45/4$).  This
improvement is significant at 97.4\% confidence.  A monochromatic
radio break of this nature is very difficult to explain physically.
However, with scintillation flux errors folded into the light curve,
we find that a simple power law is a more than adequate fit to the
data ($\chi^2/\nu = 1.4/6$), which may suggest that we have
overestimated the degree of modulation somewhat.  (This is to be
expected: the modulation index calculated is an upper limit as it
strictly applies only for a point source.  The afterglow has a
physical size, which quenches the scintillation modulation somewhat.)
Therefore, as a final modification, we scaled down this additional
error until the final $\chi^2/\nu \approx 1$.  Properties of the
temporal fits are given in Table \ref{tab:radiomodel}.

The uncertainty due to scintillation is in any event too large to
allow any firm conclusions about the light curve.  However, since only
refractive scintillation is expected to be significant, the refractive
timescale is much longer than the several-hour timescale of individual
observations, and the C-band observations were in all cases taken
immediately after the X-band observations, we do consider the
measurement of the radio spectral index ($\beta_R = -1.15 \pm 0.42$)
to be trustworthy regardless of any scintillation uncertainty.

\subsection{Spectral Energy Distribution and Extragalactic Extinction}

If our modeling assumptions are accurate (or nearly so), we can use
our model to calculate the SED at any time using a combination of all
the data available, rather than restricting the measurement to a small
subset of the photometry and filters, even if the data were acquired
at very different times in the evolution of the GRB and the color is not
constant.

We calculate the SED at two epochs.  First, we calculate the SED at $t
= 2.67$~days after the burst, the time of our four-color Gemini-South
observations.  In calculating this SED, we perform a slightly modified
light-curve fit: we do not perform any filter transformations (e.g., to
convert $r$ to $R$), but we fix all non-SED parameters to that derived
from the light-curve analysis.  In addition, we add in quadrature a
calibration uncertainty equal to 5\% in all filters, with a few
exceptions.  For $z$, we use a 15\% uncertainty.  For $u$, we use a
30\% uncertainty, for reasons described earlier.  Finally, for $K'$,
we use a large extra uncertainty of 50\% due to the possibility of a
temporal break sometime between our last optical observations and the
AO observations.  (However, if such a break is absent, then the $K'$
observation is much more precise than is given on the plots.)
Unfiltered observations are not used.  We also calculate an early-time
SED during the ``normal'' power-law decay at $t = 1000$~s, using a fit
excluding late-time measurements and measurements during the (possibly
chromatic) bump.  Addition of uncertainties is as for the late-time
SED.

\begin{figure}
\centerline{\includegraphics[width=3.5in,angle=0]{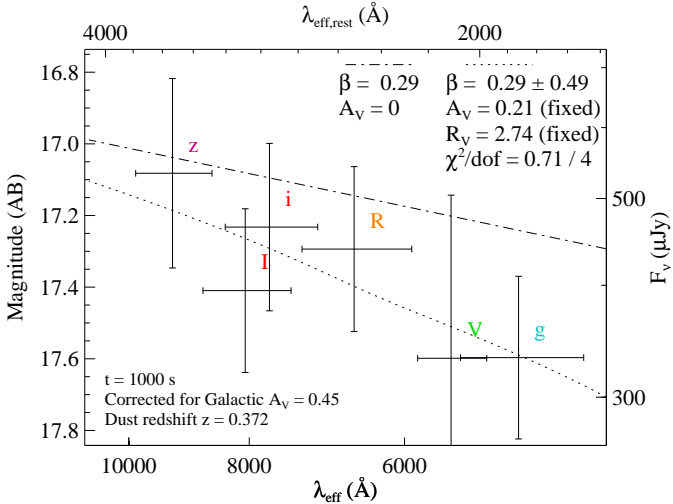}}
\caption[relation] {Optical SED of the
GRB 071003 OA at 1000~s after the burst, fitted using the extinction
constraints derived using the late-time SED.  The intrinsic
(pre-extinction) model spectrum is also shown.}
\label{fig:sed1000}
\end{figure}

\begin{figure}
\centerline{\includegraphics[width=3.5in,angle=0]{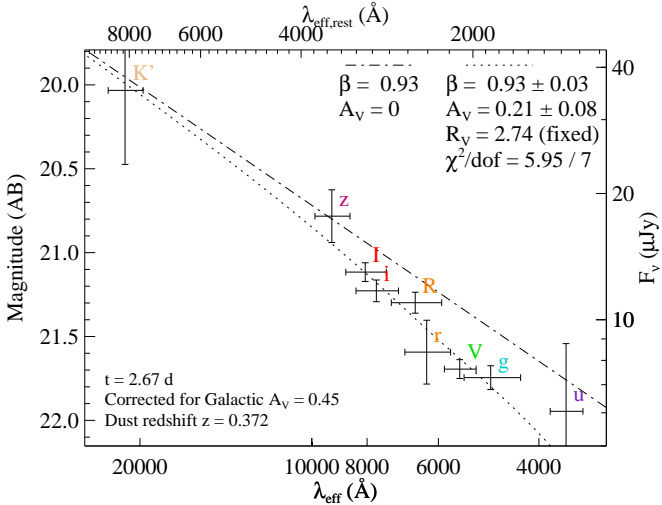}}
\caption[relation] {Same as Figure \ref{fig:sed1000} but for $t =
2.67$~d after the burst.  The data (plus an X-ray normalization,
not shown) have been fitted with an SMC-like extinction law, with the
best-fit curve overplotted.  The intrinsic (pre-extinction) model
spectrum is also shown.}
\label{fig:sed2day}
\end{figure}

The resulting SEDs are plotted in Figures \ref{fig:sed1000} and
\ref{fig:sed2day}.  After removing the effects of Galactic extinction
(but not yet considering non-Galactic extinction), both SEDs are a
reasonable fit to a power law, providing a general confirmation of our
assumptions as well as indicating that the host or intervening
galaxies do not impose a great deal of frequency-dependent extinction.
In support of our analysis from the light-curve modeling, the spectral
indices appear to differ from early to late times: $\beta_{1000s}$ =
0.62 $\pm$ 0.33, while $\beta_{2.67d}$ = 1.25 $\pm$ 0.09.  (These
values are direct fits to the data and do not include the effects of
the small amount of extragalactic extinction we do believe to be
present, which we discuss shortly.)

\begin{figure} 
\centerline{\includegraphics[width=3.5in,angle=0]{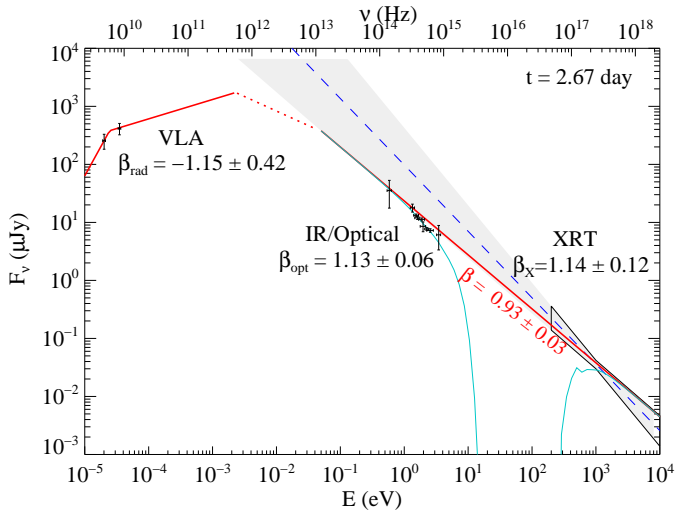}}
\caption[relation] {\small Broadband SED at
$t=2.67$~days from radio through X-ray observations.  The shaded region
shows an unbroken extrapolation of the X-ray fit (90\% confidence
region), which is consistent with the optical measurements.  The
optical points are corrected for Galactic but not extragalactic
extinction; a best-fit model for the effects of host-galaxy and
intervening-galaxy extinction is shown (thin cyan line).  The
locations of the cooling break and peak frequency shown are
arbitrarily chosen; the actual frequencies are not constrained by the
available data except that both are located between the radio and
optical bands.}
\label{fig:bbsed}
\end{figure}

Unfortunately there were no early-time observations outside the
optical band, since \textit{Swift} was unable to slew rapidly.
However, this GRB was observed nearly simultaneously in X-rays,
optical, and radio during the declining phase of the late
rebrightening.  Therefore, it is possible to calculate a coeval
late-time spectrum at all wavelengths simultaneously.  The values at
2.67~days (the same as the first optical-only SED, above, which is
also contemporaneous with XRT observations and within about half a day
of the first VLA observation) are given in Table \ref{tab:flux} and
plotted in Figure \ref{fig:bbsed}.

Even without considering host-galaxy extinction, the optical and X-ray
observations are nearly consistent with a common spectral index:
$\beta_{O}$ = 1.25 $\pm$ 0.09, $\beta_{X}$ = 1.14 $\pm$ 0.12, and
$\beta_{OX}$ = 0.90 $\pm$ 0.03.  This consistency, plus the fact that
the optical and X-ray temporal decays are identical ($\alpha_{O}$ =
1.72 $\pm$ 0.31, $\alpha_{X}$ = 1.68 $\pm$ 0.05), argues that
both X-ray and optical are in the same synchrotron regime and the
spectrum across this range is a simple power law.  We assume this
throughout the remainder of the analysis.

The deviations in the observed spectral index suggest the presence of
a small amount of extragalactic extinction.  Because of the presence
of numerous absorbers and the unusually weak nature of the
highest-redshift absorption system, however, the appropriate
assumptions for modeling the extinction contribution are not clear.
Although \ion{Mg}{2} is not an exact tracer of the presence of dust,
the extremely weak line absorption at the likely host-galaxy redshift
of $z=1.604$ suggests that the dust column at that redshift is nearly
negligible.  Among the remaining absorbers, the \ion{Mg}{2} system at
$z=0.372$ is by far the strongest (by a factor of $\sim$3 in
equivalent width compared to the next strongest system at $z=1.10$),
and is likely to be the dominant contributor to any observed dust
absorption.  However, this is partially offset by the fact that dust
at higher redshift is much more opaque (since the observed optical
frequencies are in the rest-frame UV at $z > 1$), so for the moment we
remain agnostic as to the actual redshift of the absorbing dust.

We fit the optical spectrum simultaneously with the normalized X-ray
flux of $F_{1{\rm keV}}$ = 0.036 $\pm$ 0.004 $\mu$Jy at 2.67~days.
This value has already been corrected for photoelectric absorption (\S
2.1), and X-ray absorption is not considered in the fit, allowing the
gas-to-dust ratio to be independent of the amount of extinction, $A_V$.

Four different extinction models were tested.  In addition to a
control fit with no extinction, we fit for Milky Way-like, Small
Magellanic Cloud (SMC)-like, and Large Magellanic Cloud (LMC)-like
extinction using the parameterization of the
\citet{FitzpatrickMassa1990} (``FM'') model, and a model for
extinction in starburst galaxies parameterized by
\citet{Calzetti+2000}.  In all cases the standard average value of the
ratio of total-to-selective extinction $R_V$ in the reference galaxy
in question was used.  (Fits with varying $R_V$ were attempted, but
lacking infrared or ultraviolet measurements we were unable to
constrain this parameter.)  We performed separate fits assuming dust
at $z = 0.372$, 1.100, and 1.604.

Results are given in Table \ref{tab:extfits}.  We find significant
evidence ($f$-test: 96\% confidence) for a small amount ($A_V$ =
0.1--0.3 mag, depending on the model) of extinction along the light of
sight.  We cannot strongly constrain its nature; all four extinction
laws, at each of the three possible redshifts, give reasonable fits to
the observations.  The intrinsic (pre-extinction) spectral slope
$\beta$ is strongly constrained to be 0.94 $\pm$ 0.03, averaged across
the different models.  This is consistent (although marginally, at about
the 90\% confidence level) with the absorption-corrected X-ray
measurement of $\beta$ = 1.14 $\pm$ 0.12.

As expected, the spectrum turns over dramatically somewhere redward of
the optical and is declining with decreasing frequency by the radio
band.  The radio results are discussed further in \S 5.4.

\subsection{Photometric Limits on a Host Galaxy and Intervening Absorbers}

Neither our LRIS imaging nor our late-time NGS AO imaging
show any evidence of extension or host-galaxy emission consistent with
the afterglow position.  We searched for emission from a host
coincident with the OA position by smoothing and binning the
PSF-subtracted AO image.  No host emission was detected to a
conservative upper limit of $K^\prime$ $\approx$ 23 Vega mag.

In our first-night LRIS image (when the seeing was best and
contamination from the bright nearby star relatively minimized), a
faint, extended source is visible slightly southwest of the OA.  The
same source is also visible in the AO image, clearly resolved into a
faint galaxy with $K^\prime \approx 19$ mag at an offset of
2.07$\arcsec$ southwest of the OA.

We know from the spectral analysis that there are at least four
systems that intersect the sightline between the $z=1.604$ GRB and
Earth, including the host itself.  Of these, the strongest candidate
for association with the observed galaxy is clearly the $z=0.372$
system, which both is closest and exhibits the strongest absorption
signature.  (Unfortunately, we have no spectra of the galaxy to
confirm this.)  This source appears to be a small irregular galaxy,
which at this redshift would be offset by $\sim$ 10 kpc (a reasonable
distance to explain the observed absorption) and approximately 0.5 kpc
in half-light radius.

No other extended sources are detected within 3\arcsec\ of the
afterglow, so our upper limit rules out detection of both a host
galaxy and any absorbing systems within this distance.  The
corresponding limit on a galaxy luminosity is only mild, compared to
the known GRB host distribution.  At the presumptive GRB redshift of
$z = 1.6$, any host galaxy is limited to a $K$-band absolute magnitude
of $M(K') = -22.2$ Vega mag.
This value falls roughly in the middle of the typical range of
previously studied GRB hosts, which appear to have $K$-band
luminosities on the order of 0.1 $L_*$, and are bluer and fainter than
typical SCUBA galaxies \citep{LeFloch+2003}.

\subsection{Spectroscopic Constraints on the Host Galaxy and Intervening Absorbers}

The very weak absorption at the host redshift in our spectrum suggests
a lower than average \ion{H}{1} column density along the sightline
and/or a metal-poor gas.  Because of our low spectral resolution,
however, the absorption is unresolved and the line profiles may be
saturated \citep{pro06}.  We may conservatively report a lower limit
to the column densities by assuming the weak limit.  In this manner,
we estimate $N_{Mg^+} > 10^{12.6}~ {\rm cm}^{-2}$ based on the
equivalent width of \ion{Mg}{2}~$\lambda$2803.  For a solar
metallicity gas, this implies $\log \mnhi > 10^{17}~ {\rm cm}^{-2}$.
This is a conservative estimate because the gas metallicity is
presumably subsolar.  Nevertheless, it is unlikely that the gas has an
\ion{H}{1} column density matching the values typical of most GRBs.

In addition to the gas associated with GRB~071003, the afterglow
spectrum reveals three foreground \ion{Mg}{2} absorbers.  Two of these
have moderate rest-frame equivalent widths ($W_{2796} \approx
0.7$~\AA), but the lowest redshift system exhibits a very large value
($z=0.3722$, $W_{2796} = 2.5$~\AA).  The incidence of such strong
\ion{Mg}{2} absorption at $z< 0.5$ has not yet been established along
quasar sightlines.  These absorbers are very rare at $z \approx 0.5$
however, and the incidence is declining with redshift
\citep{ntr05,ppb06}.  The number of absorbers with $W_{2796} > 1~$\AA\
per unit redshift is $\ell(z) = 0.13$ at $z = 0.5$, and the incidence
of absorbers with $W_{2796} > 2~$\AA\ is an order of magnitude lower.
This implies that one would need to observe of order 100 quasar
sightlines to detect a single absorber with $W_{2796} > 2~$\AA\ at $z<
0.5$.  Although these are {\it a posteriori} statistics, this analysis
reminds one of the apparent enhancement of strong \ion{Mg}{2}
absorbers along GRB sightlines \citep{ppc+06}.  Given its low
redshift, this system will be an excellent case to perform follow-up
observations and examine the properties of the galaxies hosting such
systems (Pollack et al. 2008, submitted)
The bright nearby star, however, poses a formidable obstacle 
for non-AO ground-based observations.

\subsection{Energetics}

The measured gamma-ray fluence of 5.32 ($-$0.67, +0.30) $\times 10^{-5}$
erg cm$^{-2}$ (Konus, 20~keV--4~MeV: \citealt{GCN6849}) can be converted
to an isotropic-equivalent total energy release in the host frame:
$E_{\rm iso}$ $=$ 3.4 ($-$0.6, +0.2) $\times 10^{53}$ erg --- well in
the upper range of \textit{Swift} events.

No clear jet break is observed over the course of our observations, in
either the optical bands or the X-ray, out to at least 6 $\times$
$10^5$~s.  There is a possible monochromatic break in the radio
bands at around 8~days ($7 \times 10^5$~s), but it appears likely
to be a scintillation artifact (see \S 4.6).

Using this limit, and following \cite{Sari+1999}, for a uniform
circumburst medium we can calculate the minimum jet opening angle and
minimum collimation-corrected energy.  Using standard values for the
radiative efficiency ($\eta = 0.5$) and circumburst density ($n = 3.0$
cm$^{-3}$) (the end result is nearly insensitive to these parameters),
we have

\begin{equation}
\theta_{\rm jet} = 6.5^{\circ} (\frac{t_{\rm jet}}{\rm d})^{3/8}
(\frac{n}{3 {\rm ~cm^{-3}}})^{1/8} (\frac{1+z}{2})^{-3/8} (\frac{E_{\rm
iso}/\eta}{10^{53} {\rm ~erg}})^{-1/8}.
\end{equation}

\noindent
However, as we discuss later, the late-time afterglow behavior in
this case favors a wind model. Thus, following \citet{LiChevalier2003}
we have

\begin{equation}
\theta_{\rm jet} = 5.4^{\circ} (\frac{t_{\rm jet}}{\rm d})^{1/4}
(A_*)^{1/4} (\frac{1+z}{2})^{-1/4} (\frac{E_{\rm iso}/\eta}{10^{53}
{\rm ~erg}})^{-1/4}
\end{equation}

The upper limit on $t_{\rm jet}$ of 7~days gives a limit on the
opening of at least 3.1 $(A_*/0.1)^{1/4}$ deg.  (As discussed
later in \S 5.4, we estimate $A_* \approx 0.1$ from the broadband
spectrum.)  Therefore the collimation-corrected energy is at least
$E_{\gamma}$ $\gtrsim$ 2 $\times 10^{50}$ $(A_*/0.1)^{1/2}$ erg.

It is also possible that the jet break is hidden by the
complicated evolution of the burst, including the rebrightening, which
would imply more modest energetics for this burst.  However, as the 
late-time slope is still relatively shallow ($\alpha = 1.72$; 
generally we expect $\alpha \geq 2$ after a jet break) we consider
this relatively unlikely.

\section {Discussion}

\subsection{Initial Power-Law Decline}

We first turn our attention to the rapidly declining power law.  The
temporal behavior of this feature is quite simple, with a decay constant
$\alpha$ = 1.466 $\pm$ 0.006 and no evident substructure before the
``bump'' or after it.  There is no evidence of a rising component or
any early break.  The observed spectral index $\beta$ = 0.62 $\pm$
0.33, although if the extinction measured at late times is also present
at early times (as we expect), the intrinsic index is actually shallower;
correcting this using our preferred extinction model, we derive $\beta$
= 0.29 $\pm$ 0.49.

Especially when the decay is observed to flatten later, very
early-time decay of this nature is often interpreted as a reverse
shock.  This seems possible --- the spectral and temporal indices are
within the range of predictions for reverse-shock models
(specifically, the thick-shell case of \citealt{Kobayashi2000}).
However, a forward-shock origin is also consistent.  Examining the
standard closure relations between $\alpha$ and $\beta$ (as in, for
example, \citealt{Price+2003}), all environment models (ISM, wind, and
jet) are consistent with the constraints derived from the data,
largely because the early-time constraint on $\beta$ is poor.  (We
discuss the forward vs.~reverse-shock models for this emission
again in \S 5.3, in connection with the late-time rebrightening.)

\subsection{The Bump: Internal Shock Origin Without a Prompt-Emission Connection}

The bump feature is of considerable interest, since it is nearly
simultaneous with a prompt-emission pulse.  However, as discussed
earlier, the temporal analysis seems to disfavor the interpretation as
a prompt reverberation: the bump seems to be already rising even
before the prompt spike.

Another possible explanation for the origin of this feature is a large
density variation in the surrounding medium (a large clump or other
discrete physical feature in the path of the expanding shock).  The
observed pulse width $\Delta t / t \approx 1$ is consistent with a
density variation, and the general appearance of the light curve over
this region is reminiscent of simulations of a GRB forward shock
intersecting ISM density enhancements (e.g., Figure 3 of
\citealt{Nakar+2003}).  However, our observation of possible color
change across the bump would (if real) disfavor this hypothesis, at least in
the simplest models: density variation will not change the intrinsic
spectrum, unless either the microphysical parameters or cooling
frequency suddenly and significantly change.  We consider this
unlikely, although some authors \citepeg{Yost+2003,Granot+2006} have
discussed the role of variable microphysics in previous GRB
afterglows.

Alternatively, the observation that the fast-declining component seems
completely unaffected by the afterglow (the temporal indices before
and after are effectively identical) leads us to interpret the bump
as originating from a distinct emission episode --- given the rapid
rise and fall and the hint of blue-to-red evolution we suggest that it
arises from internal-shock emission.  Hard-to-soft evolution and an
underlying power-law decay not affected by the flare have also been
seen in X-ray flares \citep{ButlerKocevski2007a, Chincarini+2007}. We
also note that earlier studies of GRB prompt emission have shown pulses
observed at lower energy to be broader than those at higher energy
\citep{Fenimore+1995}; this trend may continue into the optical band.
The broader, smoother profile of this pulse relative to the much
faster-evolving X-ray flares may in this case illustrate important
attributes of the emission --- either from viewing effects or resulting
from the physics of the emission itself.

\subsection{The Late Rebrightening}

The rebrightening phase of this burst is quite dramatic.  While our
observations do not sample the peak of the emission, a fit with a
reasonable assumption of the sharpness parameter suggests that the
flux increased by approximately 1 mag, and the amount of
integrated optical flux released during the rebrightening is
comparable to or more than that emitted by the early afterglow.  A
rise in optical flux of more than a magnitude at intermediate times
(well after the end of prompt emission, but before any supernova
component) has to our knowledge been seen in only a handful of 
previous cases: GRBs 970508 \citep{CastroTirado+1998}, 041219A
\citep{Blake+2005}, 060729 \citep{Grupe+2007}, 070420 \citep{GCN6338},
and 070311 \citep{Guidorzi+2007}.

The rebrightening is also notable because it appears to differ subtly
from the early decay, even though the evolution of both curves is
generally quite simple.  The decay index and spectral index both
steepen, by $\Delta\alpha$ = 0.25 $\pm$ 0.14 and $\Delta\beta$ = 0.80
$\pm$ 0.30, respectively.  Assuming a synchrotron spectrum, there are
only two possible origins for this --- the optical band is in
different synchrotron regimes at different times (specifically, $\nu <
\nu_c$ before cooling, and $\nu > \nu_c$ after cooling, consistent
with the changes observed), or because of a shift in the electron
index $p$ by approximately $\Delta p$ = +0.4.

We consider several physical origins for the rebrightening feature:
the appearance of the forward shock when the burst ejecta first
decelerate against the ISM, the late-time peak of a pre-existing
forward shock due to evolution of the critical frequencies, impact of
the forward shock through a density variation, and rebrightening
caused by a refreshed shock.

\emph{Appearance of forward shock} --- When the GRB ejecta first
begin to sweep up an amount of matter from the ISM
comparable to the energy in the ejecta, they begin to decelerate, and
reverse and forward shocks are propagated back into the ejecta and
forward into the ISM, respectively; depending on the Lorentz factor,
both shocks can then rise very quickly.  We consider this scenario
extremely unlikely to be relevant, since by necessity the forward and
reverse shocks must rise simultaneously, and there is no explanation
for the bright early-time component in the burst --- save for a prompt
model connected with internal shocks, but as we have already shown,
there is no evidence linking the early optical behavior with the
high-energy emission.

\emph{Spectral peak of existing forward shock} --- A more reasonable
model postulates that the reverse and forward shocks both formed
extremely early, but because they evolve differently (the reverse
shock, whose synchrotron parameters are boosted down by factors of
$\gamma^2$, begins to fade immediately, while the forward shock will
rise at lower frequencies), the reverse shock fades rapidly, while the
forward shock can rise and peak when the synchrotron frequency $\nu_m$
passes through the optical band.  This model has, for example, been
invoked to explain early-time bumps in the light curves of GRB~021004
\citepeg{Kobayashi+2003}, GRB~050525A \citep{ShaoDai2005}, and
GRB~080319B \citep{Bloom+2008}, which level off significantly (but do
not rebrighten) at around $10^4$ s.  However, this model is
problematic here: although we have only sparse observations of
the rebrightening, the observed rising temporal index of $\alpha$ =
$-$1.12 $\pm$ 0.16 is far too fast to be consistent with a rising
phase of a forward adiabatic shock, which predicts $F \propto
t^{(2-s)/(4-s)}$ (= $t^{1/2}$ for a constant-density ISM and constant
for a wind).  Therefore, the synchrotron peak of the forward shock
alone cannot explain this feature.

\emph{Density variation} --- A third possibility, not invoking the
transition between reverse and forward shocks, might be a dramatic
density variation: for example, the impact of the shock wave into a
previously ejected circumstellar shell, or emergence of the shock from
a low-density cavity into a dense external medium.  Density fluctuations
have been successfully invoked to explain low-level variations in
several previous studies \citepeg{Lazzati+2002} and the timescale of
the rebrightening ($\Delta t/t$ $\approx$ 1) is consistent with a
density-fluctuation origin \citep{Nakar+2003}.  However, in this case
we would expect neither a change in the spectral index (as is probably
observed) nor such a slow decline after the peak, with a temporal
index that differs significantly but only slightly from the value of
the initial decay.  Furthermore, detailed numerical studies by
numerous authors \citep{Huang+2006,Nakar+2003,Nakar+2007} have failed
to reproduce anything but the smallest rebrightening signatures in
previous GRBs using density variations.

\emph{Multi-component jet} --- The complicated light curve of GRB~030329 
has been interpreted \citep{Berger+2003} as the result of two
separate forward shocks, arising from two different jet components: a
narrow, highly relativistic jet whose emission peaks extremely early,
plus a wide, more mildly relativistic jet that dominates the late-time
and radio evolution.  Could this model conceivably explain the
observations of GRB~071003?  While a complete analysis is beyond the
scope of this paper, we note that the observations do seem consistent:
the similarities of late-time decay of both rapid and late-time
components are naturally explained, the timescale of our rebrightening
is similar to that observed in GRB~030329, and (notably) the most
significant criticism of the two-jet interpretation of GRB~030329
(that the rebrightening rose too rapidly and peaked too sharply ---
\citealt{Huang+2006}) does not apply here: the rebrightening in this
case is much smoother than that observed for GRB 030329.

\emph{Refreshed shock} --- Finally, we consider the possibility that
this feature is due to a discrete energy reinjection energizing the
forward shock, such as via a slow-moving shell that catches up to the
forward shock at late times after it decelerates.  This seems
consistent with all observations, although largely by virtue of not
making strong predictions; by invoking a customized pattern of energy
reinjection at the right times, a very broad space of light curve
behavior can be modeled \citep{Huang+2006}.  We do note that a
large, sudden rebrightening of this nature may also produce a (second)
reverse shock, which would be observable in radio and decline rapidly
with time.  The radio flux does in fact decay somewhat (in contrast to the 
expectation from a forward-shock model, where the radio flux is constant
or rising), and the measured $\alpha = 0.33 \pm 0.10$ is not far from the
predicted decay constant for a reverse shock of $\alpha \approx 1/2$ in 
the $\nu < \nu_m$ frequency regime \citep{Kobayashi2000}.  However, the
radio decay could conceivably be due to other effects (e.g., late jet break),
and without an independent measurement of the synchrotron peak frequency $\nu_m$
and late-time Lorentz factor $\Gamma$ we are unable to further constrain the 
presence or absence of such a feature with the limited observations available.

We therefore find that only the multi-component jet and refreshed
shock models are consistent with all available data.  Unfortunately,
we do not have sufficient observations during the rising phase of the
rebrightening to distinguish the two models; in particular, we can set
no constraints on the color evolution and lack a detailed light curve
of the rise to peak of the rebrightening.  We do note that the X-ray
observations are already decaying well before the (probable) optical
peak by an extrapolation of our observations (Figure \ref{fig:lc}),
which may suggest hard-to-soft evolution in this feature as well.
However, as noted earlier, the X-ray decay extrapolates back to 
the BAT light curve without explicit need for a rebrightening, so
without earlier X-ray measurements this association is speculative.

\subsection{Environmental Constraints}

In the simplest models, the late-time light curve of any GRB
is fixed by a number of basic parameters: microphysical
parameters $\epsilon_B$ (the fraction of energy in magnetic fields),
$\epsilon_e$ (the fraction of energy in electrons), and $p$ (the electron
energy index); macroscopic parameters $E_{K}$ (the blastwave
energy) and $\theta_j$ (the jet opening angle); and a parameter
quantifying the density of the surrounding medium, $n$ (for a uniform
density) or $A_*$ (for an $r^{-2}$ density profile).  Our broadband
observations (spanning from radio to X-rays) should, in principle,
allow us to firmly constrain most of these parameters for GRB~071003 ---
or, more accurately, to its late rebrightening phase, as this
component is dominant at late times.

The indices $\alpha$ and $\beta$ are both well constrained at late
times in the optical through X-ray bands, thanks to the wide range of
temporal and spatial sampling: $\alpha_{O+X}$ = 1.71 $\pm$ 0.14,
$\beta_{OX}$ = 0.93 $\pm$ 0.03.  Two environment models satisfy these
constraints within 90\% confidence: a wind-driven medium ($\rho
\propto r^{-2}$) in which $p \approx 2.9$, and a model in which the
jet break has already occurred with $p \approx 1.9$ (but consistent
with $p = 2$).  Notably, ISM models are a poor fit: the late-time
decay rate is too fast for the shallow spectral index.  The radio
observations appear to support this conclusion: the rising light curve
predicted by the ISM model is clearly ruled out, and while the slow
radio decay ($\alpha_R = 0.33 \pm 0.1$) is inconsistent in detail with
the wind prediction of constant evolution as well, it is conceivable
that variations from an exact $s=-2$ profile, an additional source of
radio emission at early times (e.g., a reverse shock), or a soft jet
break at $t \approx 5$~days may explain this difficulty.

The apparent spectral index of $\beta \approx -1.1$ observed in the
radio is notable.  A synchrotron spectrum is expected to have a
self-absorbed $\beta = -2$ spectrum below the self-absorption
frequency $\nu_a$ and a spectrum of $\beta = -0.5$ above it.  The fact
that the observed spectral index is intermediate between these values
and consistent with neither (to $\sim$90\% confidence) tells us
that, if the spectrum is really synchrotron, the absorption break is
likely to be very close to these frequencies, although exact constraints
are difficult with only two frequencies since the break is likely to be
quite soft.  The radio evolution appears nearly achromatic, which
would argue against this interpretation, but considering the
relatively narrow time and frequency window of the observations and
unknown break sharpness, we feel that this is not a major concern.

Because the ISM model is notably discrepant with the measured values
of $\alpha$ and $\beta$, we unfortunately cannot use the afterglow as
a probe of the ambient density.  If the wind model, which is more
consistent with the observations in this case, is correct, we can
calculate the parameter $A_*$ using (for example) equation 2 in
\cite{Chevalier+1999}:

\begin{equation}
F_{\nu_m} = {\rm 20~ mJy} (\frac{d_L}{\rm 5403~Mpc}) (1+z)^{1/2}
(\frac{\epsilon_B}{0.1})^{-1/2} E_{52}^{1/2} A_{*} t_{\rm d}^{1/2}.
\end{equation}

\noindent
While we have no direct measurement of $F_{\nu_m}$, it is constrained
by the radio and optical observations (see Figure \ref{fig:bbsed}) to
be $\sim$1 mJy (within a factor of $\sim$3).  We therefore
measure $A_{*} = 0.07 (\epsilon_B/0.1)^{1/2}$, an
interestingly low value regardless of the value of $\epsilon_B$.
While $\epsilon_B$ is not strongly constrained, the absence of a
cooling break between the X-ray and optical bands during the first
5 days (the cooling frequency $\nu_c$ increases in a wind model)
requires $\epsilon_B \gtrsim 0.3$.

It is possible that the wind model is inappropriate and the rapid
optical decay is due to a jet that broke before our multicolor
late-time observations.  (One possible criticism of the wind model is
that in this case, the color transition between early and late times
is hard to explain; because the cooling frequency rises with time, if
$\nu > \nu_c$ late it must have been early as well under standard
synchrotron evolution.  However, because the rebrightening appears to
be either a separate phenomenon or a large energy impulse that could
conceivably have ``reset'' the synchrotron parameters [including
$\nu_c$] to new values, this may not be a major concern.)  No jet break
is observed in the light curve, but it is possible that a jet
signature was concealed by the rebrightening.  This case would
certainly rule out the wide-angle jet interpretation of the secondary
peak and would significantly reduce the energetics.

\subsection{Spectral Implications on the Environment and Host Galaxy}

The late-time spectroscopy and imaging tell a coherent story: unlike
the vast majority of GRBs \citep{Wainwright+2007,pcw+08},
GRB~071003 did not occur in a gas-rich\footnote{Since our measurement
is based on magnesium, we are directly measuring the metal column,
not the gas column.  An alternate possibility, therefore, is that the 
host is ``normal'' but extraordinarily metal-poor, less than
$10^{-2}$ of the average solar abudance.  However, we consider a highly
subluminous host a more likely possibility.  Both effects may be
in play: low-luminosity galaxies, and those with low equivalent widths,
tend to be relatively metal-poor \citep{Prochaska+2008}.} 
galaxy.  The environment is
more consistent with a progenitor located in an outer galactic halo,
or in an extremely small (even compared to ``normal'' long-duration GRB hosts) and
gas-poor galaxy.  While the possibility of line saturation prevents us
from setting definitive upper limits, the column density through any
host is consistent with being 3 orders of magnitude below typical
GRB-derived values, and the contrast to the overall GRB population -
which is dominated by subluminous galaxies to begin with \citepeg{Fruchter+2006,Fynbo+2008},
is dramatic.

While it is well established that long-duration GRBs generally
originate from massive stars, we should be careful to ensure that our
prior experience does not blind us to the existence of rarer
subclasses of events.  We note that one other GRB on
record, GRB~070125, had very similar properties: extremely low
\ion{Mg}{2} absorption and no coincident host \citep{Cenko+2008}, as
well as a very bright afterglow and extreme energetics ($E_{\gamma}$ =
$3 \times 10^{52}$ erg; \citealt{Chandra+2007}), and even a (mild) late-time
rebrightening \citep{Updike+2008}.  Both are also among
the few \textit{Swift} bursts detected at radio wavelengths.

However, GRB~070125 and GRB~071003 show evidence from their
broadband light curves of origins typical of ordinary long GRBs.  In
the case of GRB~070125, a constant but very high circumstellar density
suggested that it occurred in what was locally a dense environment,
not an empty galactic halo, despite the near absence of a large-scale
gas signature in the spectrum.  In our case, for GRB~071003, we find
evidence of a wind-like stratified environment, a characteristic of a
massive star.  Together, these events appear to suggest an origin for
these ``halo'' bursts similar to those of all other GRBs.

If GRB~071003 did occur in a star-forming region, then there are two
possibilities consistent with the extremely small metal absorption in
the spectrum.  First, the burst may simply have formed in an extremely
subluminous galaxy --- necessarily, the number or distribution of such
objects at very high redshift is not observationally constrained, but
most simulations predict an abundance of small, highly sub-Galactic
halos in the universe that could very well harbor limited star
formation.  Alternatively, GRB~071003 may have occurred in a tidally
stripped tail from another, larger galaxy.  In this case, further
follow-up observations should reveal a disturbed, star-forming host in
the close vicinity of the burst.  

Either scenario seems plausible to explain the constraints derived on
the burst environment.  In either case, if GRBs are shown to be
reasonable tracers of star formation at high redshift, then future
large-sample GRB spectroscopy missions may be able to place important
constraints on the star-formation history of the universe not possible
by any other means.  While the sample size of such low-column-density GRBs
is now small (two events, with
GRB~061021 [\citealt{Thoene06_GCN5747}] possibly constituting a third
example), these results are already suggestive that this
fraction may be significant (on the order of a few percent), and
systematic rapid afterglow spectroscopy should continue to increase
the number considerably over the years and decades to come.  It would
be an interesting discovery if the distribution of \ion{Mg}{2}
equivalent widths turns out to be bimodal.

On a related note, the existence of GRB~071003 and GRB~070125 may have
important implications regarding the escape fraction of ionizing photons
and the reionization history of the universe.  Although the relatively
low redshift of these systems keeps the Lyman-$\alpha$ and Lyman-break
absorption features out of our spectral range and prevents us from
measuring the H~I column density directly \citep{Chen+2007}, these GRBs
provide evidence that massive stars can form well outside of gas-dense
hosts, where there is little to shield the intergalactic medium from
their ionizing UV radiation.  If the fraction of these events is more
than a few percent at $z > 7$, then such ``halo'' stars may in fact be
primarily responsible for the reionization of the universe.
Observationally, spectroscopy of such events at these high redshifts
may allow accurate measurement of the neutral gas fraction
$\bar{x}_{\rm H}$ \citepeg{McQuinn+2007} without the interference of
saturated line profiles originating from the host galaxy.

\section {Conclusions}

Although the temporal evolution of the optical afterglow of GRB 071003
is complicated, our early through late-time photometric follow-up data
clearly resolve the optical light curve into separate components.
Observations from KAIT during the prompt phase of the GRB revealed a
slowly rising, slowly falling bump or flare component, superimposed
on a simple fading power law that has no observable correlation
with the prompt emission, suggesting that while early internal-shock
flares can be observed in the optical, they are not necessarily the
same as those producing the high-energy signatures.  Our late-time
observations revealed one of the most dramatic late rebrightenings
ever recorded in a GRB light curve, and suggest that this feature is
not due to a reverse-forward shock transition or density variation,
requiring either angular jet structure or very discrete late-time
re-energizing of the optical afterglow.  This may have important
implications for the interpretation of other, less dramatic bumps and
rebrightenings at similar timescales that appear to be common features
in GRB afterglows.

The spectroscopic study of GRB 071003 offers a cautionary tale about
the standard use of \ion{Mg}{2} to infer a redshift: while it is
common practice to use the highest-redshift \ion{Mg}{2} system
observed (especially in the cases when the absorption is quite strong)
under the assumption that the GRB host system should show significant
metal absorption, here we have a clear case where this assumption is
fundamentally flawed.  Were the S/N of the spectrum worse, or the
host-galaxy absorption even weaker by a factor of only 2--3,
it is likely that we would have missed the higher-redshift system
entirely and proceeded with the assumption that this burst was at a
redshift of 1.100 instead of 1.604.  In light of this fact, previous 
and future GRB redshift claims based solely on identification of \ion{Mg}{2}
absorption should be regarded with increased skepticism.

The intervening absorption systems are nevertheless also remarkable.
With three completely independent \ion{Mg}{2} systems along the line
of sight, GRB~071003 is among the most dramatic examples yet of the
bizarre overabundance of these systems in GRB afterglows relative to
those of quasars.  Further study of this sightline, especially using
AO systems, may help shed light on this mysterious result.

\acknowledgements

KAIT and its ongoing operation were made possible by donations from
Sun Microsystems, Inc., the Hewlett-Packard Company, AutoScope
Corporation, Lick Observatory, the National Science Foundation, the
University of California, the Sylvia \& Jim Katzman Foundation, and
the TABASGO Foundation.  J.S.B.'s group is supported in part by the
Hellman Faculty Fund, Las Cumbres Observatory Global Telescope
Network, and NASA/\textit{Swift} Guest Investigator grant NNG05GF55G.
A.V.F.'s group is supported by NSF grant AST--0607485 and the TABASGO
Foundation, as well as by NASA/\textit{Swift} Guest Investigator
grants NNG05GF35G and NNG06GI86G.  N.R.B. is partially supported by a
SciDAC grant from the Department of Energy.  J.X.P. is partially
supported by NASA/\textit{Swift} Guest Investigator grant NNG05GF55G
and NSF CAREER grant AST--0548180.  H. Swan has been supported by NSF 
grant AST-0335588 and by the Michigan Space Grant Consortium. F. Yuan 
has been supported under NASA/Swift Guest Investigator Grant NNX-07AF02G. 
We acknowledge helpful discussions
with E. Ramirez-Ruiz and thank D. Whalen and A. Heger for their
calculations of the photon flux from massive stars.  We offer
particular thanks to D. A. Kann for useful discussions and feedback.

 This research is based in part on observations obtained at the Gemini
Observatory, which is operated by the Association of Universities for
Research in Astronomy, Inc., under a cooperative agreement with the
NSF on behalf of the Gemini partnership.  Some of the data presented
herein were obtained at the W.\ M.\ Keck Observatory, which is
operated as a scientific partnership among the California Institute of
Technology, the University of California, and NASA; the Observatory
was made possible by the generous financial support of the W.\ M.\
Keck Foundation. We wish to extend special thanks to those of Hawaiian
ancestry on whose sacred mountain we are privileged to be guests.
We are grateful to the staffs at the Gemini, Keck, and Lick
Observatories for their assistance.

\bibliographystyle{apj}

\clearpage

\renewcommand{\arraystretch}{0.60} 

\begin{deluxetable}{rrrrrrrrrrr}
\tabletypesize{\footnotesize}
\tablewidth{0pc}
\tablecaption{Photometry of comparison stars in the field of GRB 071003}
\tablehead{
\colhead{ID}&\colhead{$\alpha_{J2000}$\tablenotemark{a}}&\colhead{$\delta_{J2000}$\tablenotemark{a}}&\colhead{$B$}& \colhead{N$_B$} &
\colhead{$V$}&\colhead{N$_V$} & \colhead{$R$}& \colhead{N$_R$} &
\colhead{$I$} &\colhead{N$_I$}
}
\startdata 
  1&301.9066&10.9743&16.760(014)& 4&15.749(012)& 4&15.235(008)& 4&14.720(007)& 4\\
  2&301.9009&10.9357&13.760(009)& 9&12.789(007)& 8&12.251(008)& 9&11.715(009)& 8\\
  3&301.8926&10.9249&14.865(008)& 9&13.243(006)& 8&12.337(008)& 9&11.451(008)& 7\\
  4&301.8903&10.9968&15.686(006)& 6&14.748(009)& 8&14.255(009)& 9&13.818(008)& 8\\
  5&301.8811&10.9343&15.026(009)& 8&13.514(008)& 9&12.679(007)&10&11.882(006)& 7\\
  6&301.8799&10.9803&17.088(009)& 7&16.050(009)& 8&15.499(008)& 8&14.939(010)& 6\\
  7&301.8727&10.9562&17.184(014)& 4&16.372(009)& 8&15.898(008)& 8&15.479(010)& 7\\
  8&301.8705&11.0005&16.124(010)& 5&14.787(012)& 6&14.044(008)& 7&13.353(008)& 6\\
  9&301.8693&10.9253&16.473(009)& 8&15.631(008)&10&15.149(006)&10&14.714(008)& 9\\
 10&301.8688&10.9989&15.139(010)& 5&14.523(009)& 6&14.153(008)& 7&13.804(008)& 6\\
 11&301.8671&10.9417&15.310(009)& 8&14.537(008)&10&14.095(007)&10&13.692(008)&10\\
 12&301.8570&11.0033&15.828(012)& 5&15.004(012)& 4&14.569(009)& 6&14.158(010)& 4\\
 13&301.8467&10.9749&13.470(008)& 7&12.899(008)& 9&12.554(008)&10&12.220(007)& 8\\
 14&301.8396&10.9967&16.468(009)& 7&15.752(008)& 9&15.331(007)&10&14.936(009)& 9\\
 15&301.8386&10.9736&16.778(009)& 8&15.891(009)& 6&15.367(008)& 9&14.842(008)& 7\\
 16&301.8383&10.9800&15.305(010)& 7&14.034(008)& 9&13.357(006)&10&12.773(008)& 9\\
 17&301.8354&11.0043&17.455(017)& 3&16.560(004)& 3&15.970(010)& 5&15.469(008)& 5\\
 18&301.8342&10.9820&16.109(008)& 9&14.992(009)& 7&14.385(007)&10&13.832(010)& 8\\
 19&301.8334&10.9963&16.472(012)& 6&15.430(010)& 7&14.786(008)&10&14.254(008)& 9\\
 20&301.8311&11.0048&16.227(012)& 4&15.529(009)& 3&15.056(007)& 6&14.631(006)& 5\\
 21&301.8279&10.9884&17.045(008)& 7&16.107(008)& 9&15.546(008)&10&15.007(009)& 8\\
 22&301.8247&10.9781&14.276(008)& 9&13.825(008)& 9&13.510(008)&10&13.223(008)& 9\\
 23&301.8221&10.9256&16.606(009)& 9&16.002(012)& 6&15.623(007)& 9&15.204(010)& 7\\
\enddata
\tablecomments{Uncertainties (standard deviation of the mean) are 
indicated in parentheses.  This table has been truncated; additional standard stars 
are available in the online material.}
\tablenotetext{a}{In degrees.}
\label{tab:calib}
\end{deluxetable} 

\begin{deluxetable}{lrccl}
\tablecaption{KAIT photometry of GRB 071003}
\tablehead{
\colhead{$t_{\rm start}$\tablenotemark{a}}&
\colhead{Exp. time}&\colhead{Mag}&
\colhead{Error} & \colhead{Filter} \\
\colhead{(s)}&
\colhead{(s)}&\colhead{}&
\colhead{} & \colhead{}
}
\startdata 
42.0 & 5 & 12.791 & 0.019 & $R$\tablenotemark{b}  \\
49.0 & 5 & 12.999 & 0.024 & $R$  \\
55.0 & 5 & 13.193 & 0.021 & $R$  \\
61.0 & 5 & 13.321 & 0.024 & $R$  \\
67.0 & 5 & 13.500 & 0.019 & $R$  \\
97.0 & 20 & 14.465 & 0.027 &  $V$\\
128.0 & 20 & 13.919 & 0.032 & $I$\\
157.0 & 20 & 14.382 & 0.031 & $R$  \\
188.0 & 20 & 14.916 & 0.034 & $V$\\
219.0 & 20 & 14.121 & 0.022 & $I$\\
249.0 & 20 & 14.750 & 0.023 & $R$  \\
279.0 & 20 & 15.409 & 0.035 & $V$\\
310.0 & 20 & 14.578 & 0.030 & $I$\\
340.0 & 20 & 15.401 & 0.024 & $R$  \\
370.0 & 20 & 16.034 & 0.107 & $V$\\
401.0 & 20 & 15.478 & 0.063 & $I$\\
431.0 & 20 & 16.239 & 0.082 & $R$  \\
462.0 & 20 & 16.853 & 0.200 & $V$\\
492.0 & 20 & 15.977 & 0.069 & $I$\\
522.0 & 20 & 16.749 & 0.091 & $R$  \\
565.0 & 20 & 16.255 & 0.077 & $I$\\
595.0 & 20 & 16.849 & 0.097 & $R$  \\
624.0 & 20 & 16.364 & 0.106 & $I$\\
654.0 & 20 & 17.041 & 0.089 & $R$  \\
749.09\tablenotemark{c} & 3$\times$20 & 16.830 & 0.113 & $I$\\
787.68\tablenotemark{c}  & 3$\times$20 & 17.362 & 0.121 & $R$  \\
1007.74\tablenotemark{c}  & 6$\times$20 & 17.314 & 0.148 & $I$\\
1009.95\tablenotemark{c}  & 5$\times$20 & 17.711 & 0.147 & $R$  \\
1422.34\tablenotemark{c}  & 8$\times$20 & 18.103 & 0.154 & $R$ \\
1464.58\tablenotemark{c}  & 7$\times$20 & 17.473 & 0.135 & $I$\\
\enddata
\tablenotetext{a}{The start time of the exposure, in seconds after the BAT trigger.}
\tablenotetext{b}{The R-band photometry is derived from unfiltered observations.}
\tablenotetext{c}{The time (s) at the middle point of several combined images.} 
\label{tab:kait}
\end{deluxetable} 

\clearpage

\begin{deluxetable}{lcc}
\tablecaption{AEOS $R$-band photometry from unfiltered observations \tablenotemark{a}}
\tablehead{
\colhead{$t_{\rm mid}$\tablenotemark{b}}
&\colhead{$R$}&
\colhead{$\sigma_R$} \\
 \\
\colhead{(s)}&
\colhead{}& \colhead{}
}
\startdata 
568.6&16.708&0.016\\
681.3&17.046&0.018\\
794.0&17.337&0.020\\
906.8&17.573&0.020\\
1019.6&17.766&0.020\\
1132.3&17.940&0.020\\
1245.1&18.101&0.020\\
1357.9&18.229&0.023\\
1470.7&18.339&0.025\\
1583.5&18.454&0.026\\
1696.3&18.545&0.028\\
1809.1&18.640&0.034\\
1922.0&18.724&0.033\\
2034.8&18.814&0.040\\
2147.6&18.865&0.037\\
2260.3&18.924&0.042\\
2373.1&18.941&0.048\\
2485.9&18.941&0.064\\
2598.7&19.042&0.044\\
2711.4&19.087&0.049\\
2824.2&19.104&0.050\\
2937.0&19.109&0.054\\
3049.7&19.142&0.054\\
3162.5&19.150&0.052\\
3275.3&19.160&0.056\\
3388.2&19.169&0.055\\
3501.0&19.158&0.050\\
3613.8&19.139&0.056\\
3726.5&19.185&0.055\\
3839.3&19.206&0.056\\
4062.4&19.188&0.056\\
4175.3&19.196&0.055\\
4288.1&19.194&0.057\\
4401.0&19.208&0.051\\
4513.8&19.167&0.054\\
4626.6&19.150&0.053\\
4739.4&19.163&0.065\\
4852.2&19.160&0.057\\
5002.6&19.145&0.050\\
\enddata
\tablenotetext{a}{The original data set was grouped and combined into 39 images.}
\tablenotetext{b}{The time at the middle point of several combined images. }
\label{tab:aeos}
\end{deluxetable}

\begin{deluxetable}{lrccll}
\tablecaption{Keck/Gemini-S  photometry of GRB 071003}
\tablehead{
\colhead{$t_{\rm mid}$\tablenotemark{a}}&
\colhead{Exp. time}&\colhead{Mag}&
\colhead{Error} & \colhead{Filter} &\colhead{Telescope} \\
\colhead{(s)}&
\colhead{(s)}& \colhead{}&
\colhead{} & \colhead{} & \colhead{}
}
\startdata 
9523.7 & 2$\times$20 & 18.59\tablenotemark{b} & 0.25 & $C_R$ & Keck I  \\
76891.8 & 300 & 20.32 & 0.07 & $g$ & Keck I \\
77044.0 & 300 & 19.43 & 0.06 & $R$ & Keck I \\
231174.7 & 450 & 22.33 & 0.20 & $g$ & Gemini-S \\
231802.3 & 450 & 21.57 & 0.32 & $i$ & Gemini-S \\
232430.8 & 450 & 21.97 & 0.22 & $r$ & Gemini-S \\
233056.5 & 450 & 21.35 & 0.40 & $z$ & Gemini-S \\
515855.0 & 1485 & 22.61 & 0.30 & $R$ & Keck I \\
516250.5 & 975  & 23.56 & 0.30 & $g$ & Keck I \\
517510.4 & 720  & 23.56 & 0.45 & $u$ & Keck I \\
604978.0 & 660  & 23.06 & 0.50 & $R$ & Keck I \\
605144.4 & 780  & 24.05 & 0.40 & $g$ & Keck I \\
682211.1 & 330  & 23.56 & 0.40 & $V$ & Keck I \\
682211.9 & 660  & 24.42 & 0.50 & $u$ & Keck I \\
682940.0 & 840  & 23.40 & 0.60 & $R$ & Keck I \\
1373589. & 1800   & 21.58 & 0.03 & $K'$& Keck II \\
\enddata
\tablenotetext{a}{The time (s) at the middle point of the observations.} 
\tablenotetext{b}{Measured from unfiltered images from the Keck I HIRES guider.  } 
\label{tab:keck}
\end{deluxetable} 

\begin{deluxetable}{lrccl}
\tablecaption{P60 photometry of GRB 071003}
\tablehead{
\colhead{$t_{\rm start}$\tablenotemark{a}}&
\colhead{Exp. time}&\colhead{Mag}&
\colhead{Error} & \colhead{Filter} \\
\colhead{(s)}&
\colhead{(s)}&\colhead{}&
\colhead{} & \colhead{}
}
\startdata 
176.0 & 60 & 14.57 & 0.06 & $R$\\
261.0 & 60 & 15.08 & 0.05 & $i^\prime$\\
347.0 & 60 & 15.37 & 0.07 & $z^\prime$\\
432.0 & 60 & 16.03 & 0.07 & $R$\\
518.0 & 60 & 16.41 & 0.07 & $i^\prime$\\
603.0 & 60 & 16.43 & 0.08 & $z^\prime$\\
689.0 & 60 & 16.88 & 0.09 & $R$\\
775.0 & 60 & 17.12 & 0.10 & $i^\prime$\\
860.0 & 60 & 17.02 & 0.11 & $z^\prime$\\
1309.0 & 120 & 17.92 & 0.20 & $z^\prime$\\
1454.0 & 120 & 18.71 & 0.10 & $g$\\
1891.0 & 120 & 18.67 & 0.28 & $z^\prime$\\
2037.0 & 120 & 19.33 & 0.15 & $g$\\
2618.0 & 120 & 19.44 & 0.20 & $g$\\
\enddata
\tablenotetext{a}{The start time of the exposure, in seconds after the BAT trigger.}
\label{tab:p60}
\end{deluxetable} 


\begin{deluxetable}{lrrrrr}
\tablecaption{Radio observations of GRB 071003}
\tablewidth{0pt}
\tablehead{
\colhead{UT Date} & \colhead{$t_{\rm mid}$} & \colhead{Frequency} & \colhead{Flux density} & 
\colhead{Error} \\
\colhead{Observation} & \colhead{hr} & \colhead{GHz} & \colhead{$\mu$Jy} & 
\colhead{$\mu$Jy} 
}
\startdata
2007 Oct. 05, 1.85 &  42.168 & 8.46 & 393 & 55\\
2007 Oct. 07, 3.38 &  91.698 & 8.46 & 430 & 50\\
2007 Oct. 07, 3.92 &  92.238 & 4.86 & 220 & 54\\
2007 Oct. 12, 1.03 & 209.248 & 8.46 & 431 & 51\\
2007 Oct. 14, 14.84& 271.158 & 8.46 & 332 & 67\\
2007 Oct. 24, 23.58& 519.898 & 8.46 & 260 & 42\\
2007 Oct. 25, 0.04 & 520.358 & 4.86 & 119 & 46\\
2007 Nov. 05, 0.01 & 785.328 & 8.46 & 109 & 45\\
2007 Nov. 07, 0.18 & 833.336 & 4.86 & 93 & 52\\
\enddata
\label{tab:vla}
\end{deluxetable}


\begin{deluxetable}{ccccccccc}
\tablewidth{0pc}
\tablecaption{Absorption Lines in the Afterglow Spectrum of GRB~071003\label{tab:ew}}
\tabletypesize{\footnotesize}
\tablehead{\colhead{$\lambda$} & \colhead{$z$} & \colhead{Transition} & \colhead{$W^a$} & \colhead{$\sigma(W)^b$} \\
(\AA) & & & (\AA) & (\AA) }
\startdata
3549.69 & 0.37223&FeII 2586&$<2.51$&\\
3568.06 & 0.37223&FeII 2600&$2.33$&0.59\\
3837.72 & 0.37223&\ion{Mg}{2} 2796&$2.48$&0.20\\
3847.65 & 0.37223&\ion{Mg}{2} 2803&$2.14$&0.19\\
3915.45 & 0.37223&MgI 2852&$1.02$&0.17\\
4032.63 & 1.60435&CIV 1548&$0.22$&0.06\\
4039.88 & 1.60435&CIV 1550&$<0.28$&\\
4351.92 & 1.60435&AlII 1670&$<0.14$&\\
5003.26 & 1.10019&FeII 2382&$0.20$&0.05\\
5276.54 & 1.60435&ZnII 2026&$<0.08$&\\
5399.79 & 0.37223&CaII 3934&$0.61$&0.07\\
5417.99 & 0.93740&\ion{Mg}{2} 2796&$0.61$&0.05\\
5432.79$^c$ & 1.10019&FeII 2586&$0.46$&0.05\\
5447.85 & 0.37223&CaII 3969&$0.46$&0.07\\
5872.31 & 1.10019&\ion{Mg}{2} 2796&$0.80$&0.05\\
5888.27 & 1.10019&\ion{Mg}{2} 2803&$0.68$&0.06\\
6105.90 & 1.60435&FeII 2344&$<0.17$&\\
6206.91 & 1.60435&FeII 2382&$0.26$&0.04\\
6240.46 & 1.60435&FeII* 2396a&$0.25$&0.04\\
6265.95 & 1.60435&FeII* 2405&$<0.16$&\\
6282.68 & 1.60435&FeII* 2411b&$0.18$&0.03\\
6284.57 & &&$0.72$&0.12\\
6734.47 & &&$0.97$&0.15\\
6737.28 & 1.60435&FeII 2586&$0.16$&0.04\\
6772.60 & 1.60435&FeII 2600&$0.27$&0.05\\
7301.58 & 1.60435&\ion{Mg}{2} 2803&$0.17$&0.05\\
7430.06 & 1.60435&MgI 2852&$<0.24$&\\
8091.56 & &&$0.92$&0.13\\
8436.10$^d$ & &&$0.86$&0.26\\
8534.91$^d$ & &&$0.72$&0.17\\
8599.02$^d$ & &&$1.34$&0.17\\
\enddata
\tablenotetext{a}{Equivalent widths are rest-frame values and assume the redshift given in Column 2.}
\tablenotetext{b}{Limits are $2 \sigma$ statistical values.}
\tablenotetext{c}{Blended with \ion{Mg}{2}~$\lambda$2803 at $z=0.937$.}
\tablenotetext{d}{These features may be residuals from sky subtraction.}
\label{tab:lines}
\end{deluxetable}

\begin{deluxetable}{llllll}
\tablewidth{0pc}
\tablecaption{Optical Light-Curve Fits: Color Change}
\tablehead{ \colhead{Model Description} & \colhead{$\Delta\beta_{0-1}$} &  \colhead{$\Delta\beta_1(b-a)$} & \colhead{$\Delta\beta_{0-2}$} & \colhead{$\beta_2$}  & \colhead{$\chi^2/\nu$}}
\startdata
Fully monochromatic         & 0             & 0             & 0             & 0.72$\pm$0.10 & 125.765 / 81 \\
Uniformly chromatic bump    & 0.22$\pm$0.27 & 0             & 0	            & 0.68$\pm$0.10 & 125.100 / 80 \\
Variably chromatic bump     & 0.66$\pm$0.33 & 1.05$\pm$0.47 & 0             & 0.70$\pm$0.10 & 120.339 / 79 \\
Chromatic rebrightening     & 0	            & 0             & 0.77$\pm$0.31 & 1.26$\pm$0.11 & 120.040 / 80 \\  
Chromatic bump+rebrightening& 0.75$\pm$0.33 & 1.09$\pm$0.47 & 0.84$\pm$0.31 & 1.26$\pm$0.11 & 113.713 / 78 \\
\enddata
\tablecomments{Summary of relevant parameters and $\chi^2$ for models
allowing or disallowing color transitions and chromatic breaks between
the various components.  Values without uncertainties are fixed.
Component 0 is the fast-decay component, Component 1 is the bump, and
Component 2 is the late rebrightening.  The absolute late-time
spectral index $\beta_2$ is not a model parameter, but is fit
externally after completion of the fit.}
\label{tab:colorchange}
\end{deluxetable} 

\begin{deluxetable}{lllll}
\tablewidth{0pc}
\tablecaption{Optical Light-Curve Fits: $t_0$}
\tablehead{ \colhead{Model Description} & \colhead{$dt_0$} & \colhead{$dt_1$} & \colhead{$dt_2$} & \colhead{$\chi^2/\nu$} \\
            \colhead{}             & \colhead{(s)} & \colhead{(s)} & \colhead{(s)} & \colhead{}
}
\startdata
Reference          & 0             & 0             & 0            &  113.713 / 78 \\
Decay              & $-0.01 \pm 3.01$ & 0          & 0            &  113.713 / 77 \\
Bump               & 0             & 60.5$\pm$20.4 & 0            &  112.700 / 77 \\
Bump (prompt pulse)& 0             & 125.0         & 0            &  115.118 / 78 \\  
Rebrightening      & 0             & 0             & 1245$\pm$311 &  111.149 / 77 \\
\enddata
\tablecomments{Summary of relevant parameters and $\chi^2$ for models
using a $t_0$ different from the trigger time.  In all cases, the
favored color-change model (chromatic bump and rebrightening) was
used.  Values without uncertainties are fixed.  Component 0 is the
fast-decay component, Component 1 is the bump, and Component 2 is the
late rebrightening.}
\label{tab:t0}
\end{deluxetable}

\begin{deluxetable}{llll}
\tablecaption{Radio Modeling of GRB 071003}
\tablewidth{0pt}
\tablehead{
\colhead{Parameter} & \colhead{Value (broken power law)} & \colhead{Value (unbroken)} & \colhead{Value (unbroken w/scintillation\tablenotemark{a})}
}
\startdata
$\alpha_b$  & $-$0.11 $\pm$ 0.21  &  0.27 $\pm$ 0.06  & 0.34  $\pm$ 0.10 \\
$\alpha_a$  &  0.81 $\pm$ 0.25  & --                &  --\\
$t_{break}$ &  8.51 $\pm$ 3.78  & --                &  --\\
$\beta$     & $-$1.11 $\pm$ 0.34  & $-$1.15 $\pm$ 0.44  & $-$1.15 $\pm$ 0.42 \\
$\chi^2/\nu$&  2.45 / 4         &  15.32 / 6        &  6.07 / 6 \\
\enddata
\tablecomments{Best-fit parameters of a fit to the radio afterglow of
GRB 071003 using a \cite{Beuermann+1999} broken power-law model versus
an unbroken power-law model.  The improvement for the broken power-law
fit is significant given the flux uncertainties, but due to
interstellar scintillation may be coincidental.  If a small amount of
interstellar scintillation uncertainty is added in quadrature, an
unbroken power-law fit is reasonable.}  
\tablenotetext{a}{In this model, we added a 15\% error to all X-band
points and a 22\% error to all C-band points.}
\label{tab:radiomodel}
\end{deluxetable}

\begin{deluxetable}{llll}
\tablewidth{0pc}
\tablecaption{Model fluxes at $t=2.67$ days}
\tablehead{ \colhead{Band/Filter} & \colhead{E} & \colhead{Flux} & \colhead{Uncertainty} \\
            \colhead{}           & \colhead{eV} & \colhead{$\mu$Jy} & \colhead{$\mu$Jy}}
\startdata
X-ray & 1000   & 0.036 &  0.006     \\
$u$   & 3.46   &  3.17  &  1.42  \\
$g$   & 2.55   &  4.47  &  0.30 \\
$V$   & 2.25   &  5.07  &  0.27 \\
$r$   & 1.97   &  5.97  &  1.14  \\
$R$   & 1.88   &  8.01  &  0.47 \\
$i$   & 1.61   &  9.16  &  0.56 \\
$I$   & 1.54   & 10.34   &  0.54 \\
$z$   & 1.34   & 14.74   &  2.29  \\
$K$'  & 0.584  & 33.59   &  16.8  \\
X     & 3.5e-5 & 414.6   &  91.8  \\
C     & 2.0e-5 & 256.1   &  73.9  \\
\enddata
\tablecomments{Fluxes of the afterglow interpolated to $t=2.67$~d
after the BAT trigger using all available X-ray, optical, and radio
data.  Galactic extinction ($E(B-V) = 0.148$ mag) is not accounted for;
however, the X-ray flux is corrected for photoelectric absorption.}
\label{tab:flux}
\end{deluxetable}

\begin{deluxetable}{lcccll}
\tablecaption{Extinction models for optical/X-ray fits of GRB 071003}
\tablehead{
\colhead{model}&
\colhead{$A_V$}&\colhead{$R_V$}&
\colhead{$\beta$} & \colhead{$\chi^2/\nu$}
}
\startdata 
none      & 0                 & -    & 0.913 $\pm$ 0.029 & 12.4 / 8  \\
{\bf $z=0.372$} & & & & \\
Milky Way & 0.239 $\pm$ 0.093 & 3.09 & 0.939 $\pm$ 0.028 & 5.80 / 7  \\
SMC       & 0.209 $\pm$ 0.082 & 2.74 & 0.934 $\pm$ 0.028 & 5.95 / 7  \\
LMC       & 0.256 $\pm$ 0.099 & 3.41 & 0.941 $\pm$ 0.029 & 5.87 / 7  \\
Calzetti  & 0.279 $\pm$ 0.108 & 4.05 & 0.945 $\pm$ 0.029 & 5.80 / 7  \\
{\bf $z=1.10$} & & & & \\
Milky Way & 0.133 $\pm$ 0.058 & 3.09 & 0.935 $\pm$ 0.029 & 7.16 / 7  \\
SMC       & 0.127 $\pm$ 0.052 & 2.74 & 0.935 $\pm$ 0.028 & 6.38 / 7  \\
LMC       & 0.132 $\pm$ 0.057 & 3.41 & 0.934 $\pm$ 0.028 & 7.16 / 7  \\
Calzetti  & 0.247 $\pm$ 0.095 & 4.05 & 0.957 $\pm$ 0.032 & 5.78 / 7  \\
{\bf $z=1.60$} & & & & \\
Milky Way & 0.139 $\pm$ 0.048 & 3.09 & 0.943 $\pm$ 0.028 & 3.94 / 7  \\
SMC       & 0.096 $\pm$ 0.037 & 2.74 & 0.934 $\pm$ 0.028 & 5.77 / 7  \\
LMC       & 0.131 $\pm$ 0.045 & 3.41 & 0.940 $\pm$ 0.028 & 3.98 / 7  \\
Calzetti  & 0.240 $\pm$ 0.093 & 4.05 & 0.965 $\pm$ 0.033 & 5.84 / 7  \\
\enddata
\tablecomments{Results of various fits to the contemporaneous optical
and X-ray fluxes for extinction due to either the host galaxy or the
intervening absorbers at $z=0.372$ and $z=1.10$.  A small amount of
extinction is required to accurately fit the data, but its nature is
not strongly constrained.  We adopt SMC-like extinction at $z=0.372$
in the discussion and plots based on the relative strength of the
intervening absorber at this redshift in the spectrum.}
\label{tab:extfits}
\end{deluxetable}

\end{document}